%% file: Convergence_error_ArXiv.tex
\documentclass[12pt,draftcls,onecolumn]{IEEEtran}

\usepackage{amsmath}
\usepackage{amssymb}
\usepackage{amsthm}
\usepackage{graphicx}

\graphicspath{Figures/}

\newtheorem{theorem}{Theorem}
\newtheorem{definition}{Definition}
\newtheorem{lemma}{Lemma}

\newtheorem{proposition}{Proposition}
\ifCLASSINFOpdf
\else
\fi
\begin{document}
%
\title{Minimizing Convergence Error in Multi-Agent Systems via Leader Selection: A Supermodular Optimization Approach}
%
%
%

\author{Andrew~Clark$^{\ast}$,
        Basel~Alomair$^{\dagger\ast}$,
        Linda~Bushnell$^{\ast}$\footnote{$^{1}$Corresponding author.}$^{1}$,
        and~Radha~Poovendran$^{\ast}$ \\
        $^{\ast}$Department of Electrical Engineering\\
         University of Washington, Seattle, WA 98195\\
         $^{\dagger}$Center for Cybersecurity\\
         King Abdulaziz City for Science and Technology, Riyadh, Saudi Arabia\footnote{Emails: \{awclark, alomair, lb2, rp3\}@uw.edu}}
\maketitle

\begin{abstract}
In a leader-follower multi-agent system (MAS), the leader agents act as control inputs and influence the states of the remaining follower agents.  The rate at which the follower agents converge to their desired states, as well as the errors in the follower agent states prior to convergence, are determined by the choice of leader agents.  In this paper, we study leader selection in order to minimize convergence errors experienced by the follower agents, which we define as a norm of the distance between the follower agents' intermediate states and the convex hull of the leader agent states.
By introducing a novel connection to random walks on the network graph, we show that the convergence error has an inherent supermodular structure as a function of the leader set.  Supermodularity enables development of efficient discrete optimization algorithms that directly approximate the optimal leader set, provide provable performance guarantees, and do not rely on continuous relaxations.  We formulate two leader selection problems within the supermodular optimization framework, namely, the problem of selecting a fixed number of leader agents in order to minimize the convergence error, as well as the problem of selecting the minimum-size set of leader agents to achieve a given bound on the convergence error.  We introduce algorithms for approximating the optimal solution to both problems  in static networks, dynamic networks with known topology distributions, and dynamic networks with unknown and unpredictable topology distributions.  Our approach is shown to provide significantly lower convergence errors than existing random and degree-based leader selection methods in a numerical study.
\end{abstract}

\input{Introduction}

\input{Organization}

\input{Related}
\input{Preliminaries}
\input{Problem_static}
\input{Problem_dynamic}
\input{Simulation}
\input{Conclusion}

\bibliographystyle{IEEEtran}
\bibliography{TAC_response}
\appendices
\input{Proofs}

\end{document}

%% file: Introduction.tex
\section{Introduction}
\label{sec:intro}
In a multi-agent system (MAS), distributed agents exchange information  in order to compute their internal states and perform a shared task, such as achieving  consensus~\cite{ren2005survey} or maintaining and controlling a physical formation~\cite{olfati2006flocking}.  A diverse class of systems, including unmanned vehicles \cite{lawton2003decentralized}, sensor networks~\cite{olfati2009kalman}, and social networks~\cite{ghaderi2012opinion} are modeled and designed using the MAS framework.  In a leader-follower MAS, a set of leader agents acts as external control inputs in order to steer the states of the follower agents~\cite{swaroop1996string}.
  The goal of the system is for the follower agents to converge to a desired state that is determined by the set of leader agents.


 Linear weighted averaging algorithms are widely used by MAS  in domains such as parallel computing~\cite{degroot1974reaching}, distributed control~\cite{ren2007distributed}, and networking~\cite{shah2009gossip} due to their computational efficiency, distributed operation, and robustness to network topology changes.
     In such algorithms, the leader and follower agents periodically broadcast their state information and each follower agent computes its new state as a weighted average of the state values of its neighbors.  The dynamics of the  follower agents are influenced by the leader agents  through the weighted averaging rule.
    The distributed nature of these algorithms, however, can lead to errors in either the asymptotic states of the follower agents, when the followers converge to an incorrect value, or in the intermediate states of the follower agents, prior to convergence of the algorithm.
    Both types of errors in the follower agent states impact the system performance, for example, by causing formation errors in unmanned vehicle networks~\cite{yuan2012decentralised} and inaccurate estimates in sensor networks~\cite{boyd2006randomized}.

In~\cite{jadbabaie2003coordination}, it was shown that linear weighted averaging algorithms achieve asymptotic consensus to the leader agent state in both static networks and networks with dynamic topologies, provided the network is connected.  Moreover, in~\cite{tanner2004controllability} and \cite{liu2008controllability}, the authors showed that the follower agent states are controllable from a set of leader agents if the eigenvalues of the graph Laplacian are distinct.  While these existing results imply that the follower agent states can be driven to any value asymptotically, the convergence requires an arbitrary length of time, and the follower agents' intermediate states prior to convergence may deviate significantly from their desired steady-state values.


The intermediate behavior of agents under linear weighted averaging algorithms was studied in~\cite{olshevsky2009convergence}, in which upper and lower bounds on the errors in the follower agents' intermediate states were derived for static networks without leaders.  These errors were analyzed for leader-follower systems with given leader sets based on the eigenvalues of the graph Laplacian in~\cite{rahmani2009controllability} and \cite{pasqualetti2008steering}, where it was observed that the  errors depend on the given leader set.
  Hence, an efficient, analytical approach to selecting leaders in order to minimize errors in the  follower agents' intermediate states would enable the leader agents to  steer the followers to the desired state while improving performance prior to convergence.  

In this paper, we study leader selection in order to minimize the convergence error in the intermediate states of the follower agents, defined as the $l^{p}$-norm of the distance between the follower states and the convex hull of the leader agent states.  In the special case where the leader agent states are equal, this error reduces to the $l^{p}$-norm of the difference between the follower agent states and their desired steady-state values.  We formulate two leader selection problems, namely (a) selecting a set of up to $k$ leaders in order to minimize the convergence error, and (b) selecting the minimum-size set of leaders to achieve a given bound on the convergence error. We make the following specific contributions towards addressing these problems: 
\begin{itemize}
\item We derive upper bounds on the convergence error at a given time that depend on the network topology and leader set but not on the initial agent states.  We establish the equivalence between the derived upper bounds and the time for a random walk on the network graph to travel from a given follower agent to any leader agent.
\item Using the connection to random walks, we prove that the upper bound on the convergence error is a supermodular function of the leader set. We then introduce polynomial-time leader selection algorithms for problems (a) and (b) in static networks and use supermodularity to prove that the algorithms approximate the optimal convergence errors for each problem up to a provable bound.
\item We extend our approach to dynamic networks, including networks with topologies that vary in time according to a known probability distribution, and dynamic topologies that vary with unknown and unpredictable distributions.  For dynamic topologies with unknown distributions, we prove a lower bound on the best possible convergence error that can be achieved by any leader selection algorithm, as well as an upper bound on the  error achieved by our proposed algorithm.
\item Our results are illustrated through a numerical study, which compares our supermodular optimization approach with random and degree-based leader selection in static and dynamic networks.
\end{itemize}

%% file: Organization.tex
The paper is organized as follows.  In Section \ref{sec:related}, we discuss the related work.  The system model, as well as background on submodular functions and experts algorithms, are described in Section \ref{sec:model}.  The connection between convergence error and random walks, along with our proposed leader selection algorithms for networks with static topology, are presented in Section \ref{sec:static}.  Leader selection algorithms for dynamic networks are described and analyzed in Section \ref{sec:dynamic}.  Section \ref{sec:simulation} presents our numerical study.  Section \ref{sec:conclusion} concludes the paper.

%% file: Related.tex
\section{Related Work}
\label{sec:related}
We first consider the related work on  convergence of leader-follower systems and minimizing convergence error in MAS through methods other than leader selection.  We review related work on leader selection for other criteria, including controllability and minimizing error due to link noise.  We also consider previous applications of submodular optimization techniques and review related work on the connection between MAS algorithms and Markov chains.

The asymptotic convergence of MAS algorithms has been extensively studied.  Surveys of recent results can be found in \cite{mesbahi2010graph} and \cite{cao2012overview}.  The convergence rate and errors in the intermediate follower agent states have been analyzed for static consensus networks in \cite{olshevsky2009convergence}.  Subsequent works analyzed the convergence rate in networks experiencing link failures \cite{patterson2010convergence} and quantized consensus networks~\cite{cai2012convergence}.  The connection between the convergence rate and the spectrum of the network graph has been observed in leader-follower systems~\cite{rahmani2009controllability} and networks without leaders~\cite{boyd2006randomized}.  The problem of containment, defined as guaranteeing that the follower node states converge to the convex hull of the leader node states, was studied for static networks in~\cite{ji2008containment} and dynamic networks in~\cite{notarstefano2011containment}.


Current approaches to synthesizing MAS, either with a given leader  set or without leader agents, with minimum convergence error are mainly focused on optimizing the agent dynamics.  In \cite{boyd2006randomized}, the authors minimize the convergence error by optimizing the weights that each agent assigns to each of the inputs from its neighbors in a distributed fashion. In \cite{yuan2012decentralised}, an approach for dynamically modifying the weights assigned to each agent in order to increase the rate of consensus was introduced.  In addition to optimizing the behavior of the follower agents, the inputs from the leader agents can be designed using optimal control theory in order to efficiently steer the follower agents to the desired state~\cite{ji2006leader}.  These existing methods focus on minimizing  convergence errors when the leader set is given, rather than determining which agents should act as leaders, and are therefore complementary to the approach presented in this paper.


Leader selection in multi-agent systems is an emerging area of research.  In \cite{liu2011controllability}, a polynomial-time algorithm based on graph matching was introduced in order to find a set of leaders that guarantee structural controllability of the network from the leader set.  Algorithms for selecting leaders in order to minimize the impact of noise in the agent states, based on different greedy heuristics and convex relaxations, were introduced in \cite{patterson2010leader} and \cite{fardad2011noisefree,lin2011noisecorrupted}, respectively.  In \cite{kawashima2012leader}, the authors proposed a greedy algorithm for maximizing the manipulability of the network from the leader set, defined as the ratio between the norm of the follower agents' response to a given leader input and the norm of the leader input.
Supermodular optimization methods have been used to develop leader selection algorithms for minimizing the errors due to noise~\cite{clark2012supermodular} and joint optimization of performance and controllability~\cite{clark2012controllability}. 
In the preliminary version of the present paper~\cite{clark2012leader}, a supermodular optimization approach to leader selection for minimizing convergence errors in a known, static network was proposed.  In the present paper, we introduce methods for leader selection in networks with unknown and/or time-varying topologies, which was were not considered in \cite{clark2012leader}. 

Submodular optimization has also been applied in the related context of influence maximization in social networks in~\cite{kempe2003maximizing} and \cite{borkar2010manufacturing}.  The main interest of these works, however, is in maximizing the number of agents that are influenced by the leaders in steady-state, rather than the rate of convergence.  Surveys of submodular functions and submodular optimization can be found in \cite{wolsey1999integer} and \cite{fujishige2005submodular}.

The connection between multi-agent system protocols and Markov chains was first explored in \cite{chatterjee1977towards}.  More recently, a Markov chain interpretation was used to prove asymptotic convergence of consensus in networks with dynamic topologies in \cite{tahbaz2008necessary} and \cite{cao2008reaching}.  To the best of our knowledge, however, the connection between the bounds we derive on the convergence error and the Markov process consisting of a random walk on the graph does not appear in the existing literature on MAS and Markov chains.  

%% file: Preliminaries.tex
\section{System Model and Preliminaries}
\label{sec:model}
In this section, we define the system model and  convergence error metric used in this work, and give needed preliminary results on supermodular functions.  We consider a graph abstraction of the network.  To be consistent with graph-theoretic terminology, we refer to agents as \emph{nodes} in the subsequent text, with leader agents denoted as leader nodes and follower agents denoted as follower nodes.


\subsection{System Model}
\label{subsec:model}
We consider a network of $n$ agents, indexed in the set $V = \{1,\ldots,n\}$.  An edge $(i,j)$ exists from node $i$ to node $j$ if node $j$ is within communication range of node $i$.  Let $E$ denote the set of edges.  The set of neighbors of node $i$ is defined by $N(i) \triangleq \{j: (i,j) \in E\}$.  
  The number of edges outgoing from $i$ is denoted the outdegree of $i$, while the number of edges incoming to $i$ is the indegree of $i$.
We assume that the graph $G$ is strongly connected, so that for each pair of nodes $i$ and $j$, there exists a directed path from $i$ to $j$.

Each node $i$ is assumed to have a time-varying internal state, $x_{i}(t) \in \mathbb{R}$.  Each node $j$ in the leader set, denoted $S$, maintains a constant state value equal to $x_{j}^{\ast} \in \mathbb{R}$.  For every leader node $j \in S$, the state value is initialized to $x_{j}(0) = x_{j}^{\ast}$.  The leader states may be distinct, so that $x_{j}^{\ast} \neq x_{j^{\prime}}^{\ast}$ for $j,j^{\prime} \in S$.

The  follower nodes compute their states according to the distributed  rule
\begin{equation}
\label{eq:follower_dynamics}
\dot{x}_{i}(t) = \sum_{j \in N(i)}{W_{ij}(x_{j}(t)-x_{i}(t))},
\end{equation}
where $W_{ij}$ is a nonnegative constant for a given node pair $(i,j)$.    The Laplacian matrix $L$, which will be used to derive the convergence properties of the dynamics (\ref{eq:follower_dynamics}), is defined by
\begin{displaymath}
L_{ij} = \left\{
\begin{array}{ll}
-W_{ij}, & j \in N(i), \ i \notin S \\
\sum_{j \in N(i)}{W_{ij}}, & i = j, \ i \notin S\\
0, & i \in S \\
0, & \mbox{else}
\end{array}
\right.
\end{displaymath}
By \cite{chatterjee1977towards}, for any $t > 0$, $e^{-Lt}$ is a stochastic matrix with nonzero entries.  Letting $\mathbf{x}(t) \in \mathbb{R}^{n}$ denote the vector of  node states at time $t$, equation (\ref{eq:follower_dynamics}) is written in vector form as $\dot{\mathbf{x}}(t) = -L\mathbf{x}(t)$.  The node states $\mathbf{x}(t)$ depend on the leader set, $S$.  When defining the convergence error, we represent this dependence explicitly by writing $\mathbf{x}(t,S)$ to denote the node states.  For simplicity of notation, we write $\mathbf{x}(t)$ when describing the node dynamics.  
 The following lemma describes the asymptotic behavior of the dynamics in (\ref{eq:follower_dynamics}).
\begin{lemma}[\cite{ji2008containment}]
\label{lemma:convergence}
The vector of follower agent states converges to a fixed point of $\dot{\mathbf{x}}(t) = -L\mathbf{x}(t)$, denoted $\mathbf{x}^{\ast}$.  Furthermore, for each $j \in V$, $x_{j}^{\ast} \in \mbox{co}\left(\{x_{i}^{\ast} : i \in S\}\right)$, where $\mbox{co}(\cdot)$ denotes the convex hull.
\end{lemma}
Let $A = \{x_{i}^{\ast} : i \in S\}$ and $\overline{A} = \mbox{co}(A)$.  If the state of each follower node converges to $\overline{A}$, then the system is said to achieve  containment~\cite{ji2008containment}.

\subsection{Definition of Convergence Error}
\label{subsec:containment_error}
Although Lemma \ref{lemma:convergence} implies that the nodes will asymptotically converge to the convex hull of the leader node states, the node states will deviate from their desired values at each finite time.  A family of metrics for quantifying these \emph{convergence errors} in the intermediate states is defined as follows.  We let $||\cdot||_{p}$ (where $1 \leq p < \infty$) denote the $l^{p}$-norm of a vector.
\begin{definition}
\label{def:convergence_error}
Suppose $t > 0$ and $1 \leq p < \infty$.  The convergence error  $f_{t}(S)$ for leader set $S$ is defined by
\begin{equation}
\label{eq:conv_error}
f_{t}(S) \triangleq \left(\sum_{i \in V}{\left(d(x_{i}(t,S),\overline{A})^{p}\right)}\right)^{1/p} = \left(\sum_{i \in V}{\min_{y \in \overline{A}}{\left\{|x_{i}(t,S) - y|^{p}\right\}}}\right)^{1/p}.
\end{equation}
When the leaders all have the same state, $A = \overline{A} = \{x^{\ast}\}$ and $f_{t}(S) = ||\mathbf{x}(t,S) - x^{\ast}\mathbf{1}||_{p}$.
\end{definition}

The metric $f_{t}(S)$ measures the deviation of $\mathbf{x}(t,S)$ from containment at time $t$ as a function of the leader set $S$.
  In general, $f_{t}(S)$ is not  a monotonic function of $t$.  When the graph $G$ is undirected and the matrix $W$ satisfies $W_{ij} = W_{ji}$ for all $i,j$, however, the system response, and hence the convergence error is monotonic, as described in the following lemma.
\begin{lemma}
\label{lemma:monotone}
If the weight matrix $W$ is symmetric and $t_{0} > 0$, then $f_{t}(S) \leq f_{t_{0}}(S)$ for all $S \subseteq V$ and $t \geq t_{0}$.
\end{lemma}
A proof is given in the appendix.  The value of $f_{t}(S)$ depends on the value of $t$ that is chosen. One approach to choosing $t$ is to select a set $S$ at random and then select the smallest $t$ such that $f_{t}(S) \leq \beta$, where $\beta > 0$ is a desired bound on the convergence error.  This choice of $t$ guarantees that the convergence error arising from our proposed approach will be no more than $f_{t}(S)$, and hence no more than the constraint $\beta$.  Alternatively, the \emph{total convergence error}, defined as $$w(S) \triangleq \int_{0}^{\infty}{f_{t}(S) \ dt},$$ does not depend on any fixed value of $t$.  In Section \ref{sec:static}, we show that $w(S)$ can be substituted for $f_{t}(S)$ in our algorithms while preserving the optimality guarantees that we derive.

\subsection{Supermodular Functions}
\label{subsec:supermod}
In this section, we give the definition and an example of supermodular functions, as well as a  preliminary lemma that will be used in the subsequent analysis.  
\begin{definition} [Supermodularity \cite{fujishige2005submodular}]
\label{def:supermod}
Let $V$ be a finite set.  A function $f: 2^{V} \rightarrow \mathbb{R}$ is \emph{supermodular} if for every sets $S$ and $T$ with $S \subseteq T \subseteq V$ and every $v \notin T$,
\begin{equation}
\label{eq:supermod_def}
f(S) - f(S \cup \{v\}) \geq f(T) - f(T \cup \{v\}).
\end{equation}
A function $f: 2^{V} \rightarrow \mathbb{R}$ is submodular if $(-f)$ is supermodular.
\end{definition}
An example of a supermodular function is as follows.  Let $V = \{1,\ldots,n\}$ denote a set of sensor nodes, where $S \subseteq V$ is a set to be activated.  Suppose that the nodes are deployed over a region $A$, and each sensor node $i \in V$ covers a region $A_{i} \subseteq A$.  Letting $|A|$ denote the area of region $A$, the area left uncovered by $S$, defined by $$f_{1}(S) \triangleq |A| - \left|\bigcup_{i \in S}{A_{i}}\right|$$ is supermodular as a function of $S$, since adding a node to $S$ will cover a larger additional area than adding a node to a larger set $T$.

As a property of Definition \ref{def:supermod}, a nonnegative weighted sum of supermodular functions is supermodular~\cite{fujishige2005submodular}.  In addition, the following lemma gives a method for constructing supermodular functions.
\begin{lemma}[\cite{fujishige2005submodular}]
\label{lemma:supermod_convex}
Let $f: 2^{V} \rightarrow \mathbb{R}_{\geq 0}$ be a supermodular function, and suppose that $f(S) \geq f(T)$ for all $S \subseteq T$.  Let $g$ be a nondecreasing, convex, differentiable function.  Then the composition $h = g \circ f$ is supermodular.
\end{lemma}


%% file: Problem_static.tex
\section{Leader Selection in Static Networks}
\label{sec:static}

In this section, we discuss leader selection in order to minimize the convergence error in the intermediate states of networks with static topologies.  We first derive an upper bound on the convergence error that is independent of the initial leader and follower agent states, $\mathbf{x}(0)$.  We then introduce a connection between the derived upper bound and a random walk on the network graph, which provides the critical step towards formulating the leader selection problem as supermodular optimization.
Using the upper bound on the convergence error as a cost function, we formulate two leader selection problems, namely: (i) the problem of selecting a fixed set of up to $k$ leaders to minimize the convergence error, and (ii) the problem of selecting the minimum-size leader set in order to satisfy a given bound on the convergence error.  In order to efficiently approximate the solutions to (i) and (ii), we prove that the upper bound on the convergence error is a supermodular function of $S$.  Supermodularity leads to polynomial-time algorithms for approximating (i) and (ii) up to a provable constant factor.

\subsection{Random Walk-based Upper Bound on Convergence Error}
\label{subsec:random_walk}
%
As the first step in analyzing the convergence error and establishing a connection between the convergence error and a random walk on the graph $G$, an upper bound on the convergence error $f_{t}(S)$ is given by the following theorem.
\begin{theorem}
\label{theorem:upper_bound}
Let $q$ satisfy $\frac{1}{p} + \frac{1}{q} = 1$, and suppose $||\mathbf{x}(0)||_{q} \leq K$, where $K$ is a positive constant.  Further, define $P_{t} = e^{-Lt}$ and let $e_{i}$ denote the canonical vector with a $1$ in the $i$-th entry and $0$'s elsewhere.  Then for any leader set $S$, the convergence error  satisfies
\begin{equation}
\label{eq:upper_bound_equation}
f_{t}(S) \leq K\left(\sum_{i \in V \setminus S}{\left[\sum_{j \in V \setminus S}{(e_{i}^{T}P_{t})_{j}^{p}} + \left(1 - \sum_{j \in S}{(e_{i}^{T}P_{t})_{j}}\right)^{p}\right]}\right)^{1/p}.
\end{equation}
\end{theorem}

\begin{IEEEproof}
    Define $\Pi(S) \triangleq \{\pi \in \mathbf{R}_{\geq 0}^{n} : \mathbf{1}\pi = 1, \pi_i = 0 \ \forall i \notin S\}$, so that each $\pi$ defines a convex combination.  The convergence error is defined by
    \begin{displaymath}
    f_{t}(S) = \left(\sum_{i \in V}{\left[\min_{y \in \overline{A}}{\left\{|x_{i}(t) - y|^{p}\right\}}\right]}\right)^{1/p} = \left(\sum_{i \in V}{\left[\min_{\pi \in \Pi(S)}{\left\{|e_{i}^{T}P_{t}\mathbf{x}(0) - \pi^{T}\mathbf{x}(0)|^{p}\right\}}\right]}\right)^{1/p}.
    \end{displaymath}
    Bounding the above equation using H\"{o}lder's inequality yields
   $$ f_{t}(S) \leq \left(\sum_{i \in V}{\left[\min_{\pi \in \Pi(S)}{\{||\mathbf{x}(0)||_{q}^{p}||e_{i}^{T}P_{t} - \pi^{T}||_{p}^{p}\}}\right]}\right)^{1/p}
     \leq K \left(\sum_{i \in V}{\left[\min_{\pi \in \Pi(S)}{\left\{||e_{i}^{T}P_{t} - \pi^{T}||_{p}^{p}\right\}}\right]}\right)^{1/p}.$$
     Now, suppose that a distribution $\pi_{i}^{\ast} \in \Pi(S)$ is chosen such that $\pi_{i}^{\ast}(j) \geq (e_{i}^{T}P_{t})_{j}$ for all $j \in S$.  It is always possible to construct such a distribution since $\mathbf{1}^{T}\pi = 1$ for all $\pi \in \Pi(S)$ and $\sum_{j \in S}{(e_{i}^{T}P_{t})_{j}} \leq 1$. Define $\hat{\pi}_{i}(j) = \pi^{\ast}_{i}(j) - (e_{i}^{T}P_{t})_{j}$.  Then we have
     $$f_{t}(S) \leq K \left(\sum_{i \in V \setminus S}{\left[\sum_{j \in V \setminus S}{|(e_{i}^{T}P_{t})_{j}|^{p}} + \sum_{j \in S}{\hat{\pi}_{i}(j)^{p}}\right]}\right)^{1/p}
     \leq K\left(\sum_{i \in V \setminus S}{\left[\sum_{j \in V \setminus S}{(e_{i}^{T}P_{t})_{j}^{p}} + \left(\sum_{j \in S}{\hat{\pi}_{i}(j)}\right)^{p}\right]}\right)^{1/p},$$
     where the bound on the first term follows from the fact that $\hat{\pi}_{i}(j) \geq 0$ for all $i \in V \setminus S$ and $j \in S$.  Using the facts that $\hat{\pi}_{i}(j) = \pi^{\ast}_{i}(j) - (e_{i}^{T}P_{t})_{j}$ and $\sum_{j \in S}{\pi^{\ast}_{i}(j)} = 1$ yields the desired result.
    \end{IEEEproof}
    We observe that, while the bound (\ref{eq:upper_bound_equation}) is loose in general, the left and right hand sides of (\ref{eq:upper_bound_equation}) become arbitrarily close as $t$ grows large.


We define the notation $\hat{f}_{t}(S)$ for the upper bound on the convergence error $f_{t}(S)$ as given in Theorem \ref{theorem:upper_bound}, as
\begin{equation}
\label{eq:hat_epsilon_def}
\hat{f}_{t}(S) \triangleq \sum_{i \in V \setminus S}{\left[\sum_{j \in V \setminus S}{(e_{i}^{T}P_{t})_{j}^{p}} + \left(1 - \sum_{j \in S}{(e_{i}^{T}P_{t})_{j}}\right)^{p}\right]}.
\end{equation}

We now establish a mathematical relationship between the terms of the inner summation of (\ref{eq:hat_epsilon_def}) and a random walk on $G$ that will be used to prove supermodularity of $\hat{f}_{t}(S)$ later.  Intuitively, the inputs from the leader nodes can be viewed as diffusing from the leader nodes to the follower nodes via a random walk.  The connection is described formally as follows.

We define a random walk on $G$ as follows.  Choose $\delta > 0$ such that $t = \tau \delta$ for some positive integer $\tau$.  Define $X(\tau)$ to be a random walk with transition matrix $P_{\delta} \triangleq e^{-L\delta}$ (as in Theorem \ref{theorem:upper_bound}).  The following is a standard result, which we prove for completeness. 

\begin{theorem}
\label{theorem:random_walk}
Choose $\delta > 0$ such that $t = \tau\delta$ for some integer $\tau$. Let $X(\tau)$ be a random walk on $G$ with transition matrix $P_{\delta}$.  Then
    \begin{IEEEeqnarray}{rCl}
    \label{eq:containment_rw_first}
    (e_{i}^{T}P_{\delta}^{\tau})_{j} &=& Pr(X(\tau) = j | X(0) = i),\\
    \label{eq:containment_rw_second}
    1 - \sum_{j \in S}{(e_{i}^{T}P_{\delta}^{\tau})_{j}} &=& Pr(X(\tau) \notin S | X(0) = i),
    \end{IEEEeqnarray}
    where $P_{\delta}^{\tau}$ is equal to the matrix $P_{\delta}$ raised to the $\tau$-th power.
\end{theorem}

\begin{IEEEproof}
The vector $e_{i}$ defines a probability distribution on the set of nodes $V$, corresponding to the case where $X(0) = i$.  Hence, $e_{i}^{T}P_{\delta}^{\tau}$ is the probability distribution of $X(\tau)$, conditioned on $X(0) = i$, so that $(e_{i}^{T}P_{\delta}^{\tau})_{j} = Pr(X(\tau) = j|X(0) = i)$.  Eq. (\ref{eq:containment_rw_first}) follows immediately, while $$1 - \sum_{j \in S}{(e_{i}^{T}P_{\delta})_{j}} = 1 - \sum_{j \in S}{Pr(X(\tau) = j | X(0) = i)} = 1-Pr(X(\tau) \in S) = Pr(X(\tau) \notin S)$$ yields (\ref{eq:containment_rw_second}).
\end{IEEEproof}


\subsection{Problem Formulation -- Selecting up to $k$ Leaders}
\label{subsec:static_k}
In this section, we first formulate the problem of selecting a set of up to $k$ leaders, denoted $S$, in order to minimize the convergence error bound $\hat{f}_{t}(S)$. We then prove that $\hat{f}_{t}(S)$ is supermodular as a function of $S$, leading to an efficient algorithm for approximating the optimal leader set.

Selecting up to $k$ leaders in order to minimize the convergence error bound $\hat{f}_{t}(S)$ is given by the optimization problem
\begin{equation}
\label{eq:primal_static_k}
\begin{array}{cc}
\mbox{minimize} \vspace{-0.1in}
 & \hat{f}_{t}(S) \\
 \vspace{-0.1in} S & \\
\mbox{s.t.} & |S| \leq k
\end{array}
\end{equation}

Since an exhaustive search over the feasible values of $S$ is computationally prohibitive, we investigate the structure of the convergence error bound  $\hat{f}_{t}(S)$ in order to develop an efficient algorithm for approximating the solution to (\ref{eq:primal_static_k}).  By showing that $\hat{f}_{t}(S)$ satisfies supermodularity as a function of $S$, we  derive polynomial-time algorithms with  provable $O(1)$ optimality gap.

 As a first step to proving that $\hat{f}_{t}(S)$ is supermodular,  we prove  the following intermediate result that the probability that a random walk on $G$, originating at follower node $i \in V \setminus S$, does not reach any node in the leader set is a supermodular function of $S$.  

\begin{lemma}
\label{lemma:main_result_helper}
Define $g_{ij}^{\tau}(S) \triangleq Pr(X(\tau) = j | X(0) = i)$ and $h_{i}^{\tau}(S) \triangleq Pr(X(\tau) \notin S | X(0) = i)$.  Then for all $i \in V \setminus S$, $j$, and $\tau$, $g_{ij}^{\tau}(S)$ and $h_{i}^{\tau}(S)$ are both supermodular functions of $S$.
\end{lemma}
\begin{IEEEproof}
Let $S \subseteq T$ and let $u \in V \setminus T$ (and hence $u \in V \setminus S$).  Let $A_{ij}^{\tau}(S)$ denote the event that $X(\tau) = j$ and $X(r) \notin S$ for all $1 \leq r \leq \tau$, and define $\chi(\cdot)$ to be the indicator function of an event.  Since  each node in $S$ is an absorbing state of the walk, we have
\begin{displaymath}
g_{ij}^{\tau}(S) = Pr(A_{ij}^{\tau}(S) | X_{0} = i) = \mathbf{E}(\chi(A_{ij}^{\tau}(S))|X_{0}=i).
\end{displaymath}
 Furthermore, let $B_{ij}^{\tau}(S,u)$ denote the event where $X(0)= i$, $X(\tau) = j$, $X(r) \notin S$ for $0 \leq r \leq \tau$, and $X(m) = u$ for some $0 \leq m \leq \tau$.  We then have $A_{ij}^{\tau}(S) = A_{ij}^{\tau}(S \cup \{u\}) \cup B_{ij}^{\tau}(S,u)$, $A_{ij}^{\tau}(S \cup \{u\}) \cap B_{ij}^{\tau}(S,u) = \emptyset$, and
\begin{displaymath}
\chi(A_{ij}^{\tau}(S)) = \chi(A_{ij}^{\tau}(S \cup \{u\})) + \chi(B_{ij}^{\tau}(S,u)).
\end{displaymath}
Since $S \subseteq T$, $X(r) \notin T$ for all $0 \leq r \leq \tau$ implies $X(r) \notin S$ for all $0 \leq r \leq \tau$, i.e., $B_{ij}^{\tau}(T,u) \subseteq B_{ij}^{\tau}(S,u)$.  We have
$$\chi(A_{ij}^{\tau}(S)) - \chi(A_{ij}^{\tau}(S \cup \{u\}) = \chi(B_{ij}^{\tau}(S,u)) \geq \chi(B_{ij}^{\tau}(T,u)  =  \chi(A_{ij}^{\tau}(T)) - \chi(A_{ij}^{\tau}(T \cup \{u\})).$$
Taking expectations of both sides yields
\begin{IEEEeqnarray*}{rCl}
g_{ij}^{\tau}(S) - g_{ij}^{\tau}(S \cup \{u\}) &=& \mathbf{E}(\chi(A_{ij}^{\tau}(S))) - \mathbf{E}(\chi(A_{ij}^{\tau}(S \cup \{u\}))) \\ &\geq& \mathbf{E}(\chi(A_{ij}^{\tau}(T))) - \mathbf{E}(\chi(A_{ij}^{\tau}(T \cup \{u\})))
 = g_{ij}^{\tau}(T) - g_{ij}^{\tau}(T \cup \{u\}),
 \end{IEEEeqnarray*}
implying that $g_{ij}^{\tau}(S) = \mathbf{E}(\chi(A_{ij}^{\tau}(S)))$ is supermodular as a function of $S$.
A similar argument shows that $h_{i}^{\tau}(S)$ is supermodular as a function of $S$.
\end{IEEEproof}


We can now prove that $\hat{f}_{t}(S)$ is supermodular as a function of $S$ by using Lemma \ref{lemma:main_result_helper} and the composition result of Lemma \ref{lemma:supermod_convex}.

\begin{theorem}
\label{theorem:main_result}
The convergence error bound $\hat{f}_{t}(S)$ is supermodular as a function of $S$.
\end{theorem}

\begin{IEEEproof}
By Theorem \ref{theorem:random_walk}, the convergence error bound $\hat{f}_{t}(S)$ can be written as
\begin{displaymath}
\hat{f}_{t}(S) = \sum_{i \in V \setminus S}{\left[\sum_{j \in V \setminus S}{(e_{i}^{T}P_{t})_{j}^{p}} + \left(1 - \sum_{j \in S}{(e_{i}^{T}P_{t})_{j}}\right)^{p}\right]} = \sum_{i \in V}{\sum_{j \in V \setminus S}{g_{ij}^{\tau}(S)^{p}} + h_{i}^{\tau}(S)^{p}}.
\end{displaymath}
Since $x^{p}$ is a nondecreasing, convex, and differentiable function of $x$, $g_{ij}^{\tau}(S)^{p}$ and $h_{i}^{\tau}(S)^{p}$ are supermodular functions of $S$ for all $i$, $j$, $\tau$, and $p \in [1, \infty)$ by Lemmas \ref{lemma:supermod_convex} and \ref{lemma:main_result_helper}.  Hence $\hat{f}_{t}(S)$ is a sum of supermodular functions, and is therefore supermodular.
\end{IEEEproof}
As a corollary to Theorem \ref{theorem:main_result}, the total convergence error $w(S)$ defined in Section \ref{subsec:containment_error} is the integral of a family of supermodular functions, and is therefore supermodular.  As a result, the algorithms \emph{Select-$k$-leaders} and \emph{Select-minimal-leaders} can be modified by replacing $\hat{f}_{t}(S)$ with $w(S)$.

An algorithm for approximating the optimal solution to (\ref{eq:primal_static_k}) is as follows.  Initialize the set $S = \emptyset$.  At the $j$-th iteration, choose the node $v_{j} \in V \setminus S$ such that $\hat{f}_{t}(S) - \hat{f}_{t}(S \cup \{v_{j}\})$ is maximized.  The algorithm terminates after $k$ iterations.  A pseudocode description is given as algorithm \emph{Select-$k$-leaders}.
\begin{figure}[h]
\centering
\renewcommand{\arraystretch}{0.6}
\begin{tabular}{l}
\hline
\small{\textbf{Algorithm} \textbf{Select-k-leaders}: Algorithm for selecting up to $k$ leaders to minimize convergence error bound $\hat{f}_{t}$.} \\
\small{\textbf{Input:} Number of leaders $k$, network topology $G=(V,E)$, weight matrix $W$} \\
\small{\textbf{Output:} Leader set $S$}\\
\small{\textbf{Initialization:} $S \leftarrow \emptyset$, $j \leftarrow 0$} \\
\small{\textbf{while}($j < k$)} \\
\small{~~$v_{j} \leftarrow \arg\max{\{\hat{f}_{t}(S) - \hat{f}_{t}(S \cup \{v\}) : v \in V \setminus S\}}$} \\
\small{~~$S \leftarrow S \cup \{v_{j}\}$}, \small{~~$j \leftarrow j+1$} \\
\small{\textbf{end while}} \\
\small{\textbf{return} $S$} \\
\hline
\end{tabular}
\end{figure}

The following theorem gives a worst-case bound on the optimality gap between the best possible solution to (\ref{eq:primal_static_k}) and the convergence error of the set $S$ returned by \emph{Select-$k$-leaders}.  The bound guarantees that the convergence error of \emph{Select-$k$-leaders} is within a constant factor of the lowest possible convergence error.
\begin{theorem}
\label{theorem:static_k_error}
Let $S^{\ast}$ denote the set of leader nodes that is the solution of (\ref{eq:primal_static_k}).  Then algorithm \emph{Select-$k$-leaders} returns a set $S^{\prime}$ satisfying
\begin{equation}
\label{eq:primal_static_bound}
\hat{f}_{t}(S^{\prime}) \leq \left(1 - \frac{1}{e}\right)\hat{f}_{t}(S^{\ast}) + \frac{1}{e}f_{max},
\end{equation}
where $f_{max} = \max_{v \in V}{\hat{f}_{t}(\{v\})}$ and $e$ is the base of natural logarithms.
\end{theorem}
\begin{IEEEproof}
Theorem 9.3 of \cite[Ch III.3.9]{wolsey1999integer} states that, for a nonnegative nondecreasing submodular function $f(S)$, a maximization algorithm that chooses
\begin{displaymath}
v_{t} = \arg\max{\{f(S \cup \{v\}) - f(S) : v \in V \setminus S\}}
  \end{displaymath}
  returns a set $S^{\prime}$ satisfying $f(S^{\prime}) \geq \left(1 - \frac{1}{e}\right)f(S)$ for all $S \subseteq V$.  Algorithm \emph{Select-$k$-leaders} is equivalent to greedy maximization of the nonnegative, nondecreasing submodular function $f_{max} - \hat{f}_{t}(S)$.  Hence
$f_{max} - \hat{f}_{t}(S^{\prime}) \geq \left(1 - \frac{1}{e}\right)(f_{max} - \hat{f}_{t}(S^{\ast}))$.
Rearranging terms gives (\ref{eq:primal_static_bound}).
\end{IEEEproof}

\subsection{Problem Formulation -- Selecting the Minimum-size Leader Set that Achieves an Error Bound}
\label{subsec:static_alpha}
In this section, we consider the problem of selecting the minimum-size leader set $S$ in order to achieve a given constraint, $\alpha \geq 0$, on the convergence error bound $\hat{f}_{t}(S)$.  We first give the problem formulation, followed by an efficient algorithm for approximating the optimal leader set.  The supermodularity of $\hat{f}_{t}(S)$ leads to a provable $O(1)$ optimality gap between the size of the selected leader set and the size of the smallest possible leader set that achieves the convergence error $\alpha$.

Selecting the minimum-size leader set that achieves a bound on the convergence error is given by the optimization problem
\begin{equation}
\label{eq:static_alpha}
\begin{array}{cc}
\mbox{minimize}\vspace{-0.1in} & |S| \\
S \vspace{-0.1in} & \\
\mbox{s.t.} & \hat{f}_{t}(S) \leq \alpha
\end{array}
\end{equation}
Since $\hat{f}_{t}(S)$ is a supermodular function of $S$ by Theorem \ref{theorem:main_result}, equation (\ref{eq:static_alpha}) is a supermodular optimization problem, which can be efficiently approximated by a greedy algorithm analogous to \emph{Select-$k$-leaders}.
The algorithm begins by initializing $S = \emptyset$.  At the $j$-th iteration, the algorithm selects the node $v_{j}$ that maximizes $(\hat{f}_{t}(S) - \hat{f}_{t}(S \cup \{v_{j}\}))$ and sets $S = S \cup \{v_{j}\}$.  A pseudocode description of the algorithm is given as \emph{Select-minimal-leaders} below.
\begin{figure}[h]
\centering
\renewcommand{\arraystretch}{0.6}
\begin{tabular}{l}
\hline
\small{\textbf{Algorithm Select-minimal-leaders}: Algorithm for selecting the minimum-size leader set $S$ such that $\hat{f}_{t}(S) \leq \alpha$.} \\
\small{\textbf{Input:} Convergence error bound $\alpha$, network topology $G=(V,E)$, weight matrix $W$} \\
\small{\textbf{Output:} Leader set $S$} \\
\small{\textbf{Initialization:} $S \leftarrow \emptyset$} \\
\small{\textbf{while}($\hat{f}_{t}(S) > \alpha$)}\\
~~\small{$v_{j} \leftarrow \arg\max{\{\hat{f}_{t}(S) - \hat{f}_{t}(S \cup \{v\}): v \in V \setminus S\}}$} \\
~~\small{$S \leftarrow S \cup \{v_{j}\}$} \\
\small{\textbf{end while}} \\
\small{\textbf{return} $S$} \\
\hline
\end{tabular}
\end{figure}

Bounds on the optimality gap of the solutions returned by \emph{Select-minimal-leaders} are given by the following theorem.

\begin{theorem}
\label{theorem:static_alpha}
Let $S^{\ast}$ be the optimum set of leaders for problem (\ref{eq:static_alpha}), and let $S^{\prime}$ be the set of leaders returned by \emph{Select-minimal-leaders}.  Then 
$\frac{|S^{\prime}|}{|S^{\ast}|} \leq 1 + \ln{\left(\frac{f_{max}}{\alpha}\right)}$.
\end{theorem}

\begin{IEEEproof}
Theorem 9.4 of \cite[Ch III.3.9]{wolsey1999integer} implies that, for any nonnegative, nondecreasing, submodular function $f(S)$, the set $S^{\prime}$ returned by the greedy maximization algorithm and the optimal set $S^{\ast}$ satisfy
\begin{displaymath}
\frac{|S^{\prime}|}{|S^{\ast}|} \leq 1 + \ln{\left\{\frac{f(V) - f(\emptyset)}{f(V) - f(S_{T-1})}\right\}},
\end{displaymath}
where $S_{T-1}$ denotes the leader set at the second-to-last iteration of \emph{Select-minimal-leaders}.
Applying this result to the submodular function $f(S) = f_{max} - \hat{f}_{t}(S)$, and using the fact that $\hat{f}_{t}(S^{\prime}) \leq \alpha$ yields
\begin{displaymath}
\frac{|S^{\prime}|}{|S^{\ast}|} \leq 1 + \ln{\left\{\frac{-f_{max}}{-\hat{f}_{t}(S_{k-1})}\right\}}
 = 1 + \ln{\left\{\frac{f_{max}}{\hat{f}_{t}(S_{k-1})}\right\}}
\leq 1 + \ln{\left\{\frac{f_{max}}{\alpha}\right\}},
\end{displaymath}
as desired.
\end{IEEEproof}

We have that, for any $l \in V$, $$\hat{f}_{t}(\{l\}) = \sum_{i \in V \setminus \{l\}}{||e_{i}^{T}P_{t} - e_{0}^{T}||_{p}^{p}} \leq \sum_{i \in V \setminus \{l\}}{||e_{i}^{T}P_{t} - e_{0}^{T}||_{1}^{p}} \leq n-1,$$ and hence $f_{max} = \max{\{\hat{f}_{t}(\{l\}) : l \in V\}} \leq n-1$.  Thus for fixed $\alpha$, the bound of Theorem \ref{theorem:static_alpha} is of $O(\ln{n})$ in the worse case.

%% file: Problem_dynamic.tex
\section{Leader Selection in Time-Varying Networks}
\label{sec:dynamic}
In this section, we consider leader selection for two cases of networks with time-varying topologies.  In the first case, we assume that the network topology evolves according to a stochastic process with known probability distribution.  Examples of this type of topology include networks that experience random link failures with known spatial correlation~\cite{jakovetic2010weight} and networks that change topology according to a known switching signal.  In the second case, the network topology evolves according to a stochastic process with distribution that is unknown to the system designer at the time when the leaders are selected, for example, if the agents move to avoid unforeseen obstacles~\cite{olfati2006flocking}.  For each case, we first define the graph model, and then present leader selection algorithms along with  bounds on the optimality gap between the convergence error guaranteed by our algorithm and the optimal leader set.   In the case of topology dynamics with unknown distribution, we also give the best-case performance of any possible algorithm without probability distribution information and prove that our approach achieves a constant factor of this performance.


\subsection{Dynamic Topology Model}
\label{subsec:dynamic_model}
We assume that the set of network nodes, denoted $V$, is constant.  The set of edges is represented by a random process $E(t)$, resulting in a time-varying network topology $G(t) = (V, E(t))$.  We assume that   there exists a sequence of random variables $t_1, t_2,\ldots$, such that $E(t) = E(t^{\prime})$ for all $t, t' \in [t_m, t_{m+1}]$, and that $\inf_{m}{|t_{m+1}-t_{m}|} = \gamma > 0$.  We assume that $G(t)$ is strongly connected for all $t \geq 0$.

The weight matrix at time $t$ is denoted $W(t)$, while the corresponding Laplacian matrix is given by $L(t)$.  The dynamics of the follower nodes are governed by
\begin{equation}
\label{eq:follower_dynamics_dynamic}
\dot{\mathbf{x}}(t) = -L(t)\mathbf{x}(t),
\end{equation}
while each leader node $i \in S$ maintains  a constant state $x_{i}^{\ast}$ and $\overline{A} = \mbox{co}\{x_{i}^{\ast} : i \in S\}$ is the convex hull of the leader node states.  Let $r$ denote the number of network topology changes up to time $t$, i.e., $r = \max{\{m: t_{m} \leq t\}}$, and let $\delta_{m} = t_{m}-t_{m-1}$. 
The agent states at time $t$ are given by
\begin{displaymath}
\mathbf{x}(t) = \left(\prod_{m=1}^{r}{e^{-L(t_{m-1})\delta_{m}}}\right)\mathbf{x}(0).
\end{displaymath}
 The following lemma describes the convergence properties of (\ref{eq:follower_dynamics_dynamic}).
\begin{lemma}[\cite{notarstefano2011containment}]
\label{lemma:dynamic_convergence}
Under the assumption that $G(t)$ is strongly connected $\forall t$, $\lim_{t \rightarrow \infty}{d(x_{i},\overline{A})} = 0$ for every node $i \in V$.
\end{lemma}

As a corollary, when the leaders have the same initial state $x^{\ast}$, the follower nodes will asymptotically converge to $x^{\ast}$. While Lemma \ref{lemma:dynamic_convergence} implies that any leader set guarantees asymptotic convergence provided that the graph $G(t)$ is strongly connected for all $t \geq 0$, asymptotic convergence alone does not imply that the errors in the intermediate states of the follower nodes are minimized.  Selecting the  leader set in order to minimize the intermediate state convergence errors is the focus of the next section.  

One example of a type of topology dynamics that can be analyzed within this framework is the \emph{random waypoint model} \cite{bettstetter2003node}.  Under this model, each node chooses a destination  uniformly at random from within a certain deployment area and moves toward that destination with a randomly chosen velocity, choosing a new destination upon arrival.  If the mobility model has reached its stationary distribution, then the expected convergence error $\hat{f}_{t}(S)$ can be bounded by
\begin{equation}
\label{eq:unknown_topology_approx}
\mathbf{E}\left\{\sum_{i \in V \setminus S}{\sum_{j \notin S}{\left(e_{i}^{T}\prod_{m=1}^{r}{e^{-L(t_{m-1})\delta_{m}}}\right)_{j}^{p} + \left(1 - \sum_{j \in S}{\left(e_{i}^{T}\prod_{m=1}^{r}{e^{-L(t_{m-1})\delta_{m}}}\right)_{j}}\right)^{p}}}\right\}.
\end{equation}
     The value of (\ref{eq:unknown_topology_approx}) can be estimated using Monte Carlo methods, yielding an approximation of the convergence error when the topology varies according to a known random waypoint mobility model.



\subsection{Leader Selection Under Known Topology Distribution}
\label{subsec:known_dynamics}
We first treat the case where, at each time $t$, the distribution of the random variable $G(t)$ is known to the system designer.  In this case, if the leader set varies over time, then selecting a leader set for each topology in order to minimize the convergence error is equivalent to solving a series of problems of the form (\ref{eq:primal_static_k}).

The case of minimizing the convergence error when the leader set is fixed is discussed as follows.
As in Section \ref{subsec:model}, we measure the convergence error by $f_{t}(S) \triangleq \left(\sum_{i \in V}{\left(d(x_{i}(t,S),\overline{A})^{p}\right)}\right)^{1/p}$.  

%

A straightforward extension of Theorem \ref{theorem:upper_bound} implies that $f_{t}(S)$ is bounded above by $\hat{f}_{t}(S)$, which is defined for time-varying networks by
\begin{equation}
\label{eq:hat_epsilon_dynamic}
\hat{f}_{t}(S) \triangleq \mathbf{E}\left\{\sum_{i \in V \setminus S}{\left[\sum_{j \in V \setminus S}{\left(e_{i}^{T}\prod_{m=1}^{r}{e^{-L(t_{m-1})\delta_{m}}}\right)_{j}^{p}} + \left(1 - \sum_{j \in S}{\left(e_{i}^{T}\prod_{m=1}^{r}{e^{-L(t_{m-1})\delta_{m}}}\right)_{j}}\right)^{p}\right]}\right\},
\end{equation}
where the expectation is over the possible realizations of $\{G(\tau): 0 \leq \tau \leq t\}$.
The following theorem leads to efficient algorithms for minimizing the convergence error for topology dynamics with known distribution.
\begin{theorem}
\label{theorem:supermodular_dynamic}
The  convergence error bound $\hat{f}_{t}(S)$ is supermodular as a function of $S$.
\end{theorem}

The proof is given in the appendix. Theorem \ref{theorem:supermodular_dynamic} implies that the algorithms \emph{Select-$k$-leaders} and \emph{Select-minimal-leaders} from Section \ref{sec:static} can be modified to select leaders in the known topology distribution case by using the generalized version of $\hat{f}_{t}(S)$ defined above.  The optimality gap provided by Theorems \ref{theorem:static_k_error} and \ref{theorem:static_alpha} are maintained in the known topology dynamics case.

\subsection{Leader Selection Under Unknown Topology Distribution}
\label{subsec:unknown}
We next consider the case where the distribution of the random process $G(t)$ is not known.  The distribution of the topology dynamics may be unknown, for example, due to node mobility, which creates and removes links at arbitrary and random points in time.  For this case, we first define an upper bound on the convergence error for topologies with unknown distribution, and then formulate the problem of selecting up to $k$ leaders in order to minimize this upper bound on the convergence error.  We then define the notion of regret, which will be used to evaluate our proposed leader selection approach, and give a lower bound on the regret that can be achieved by any algorithm.    We give background on expert algorithms that are called as subroutines of our approach, and finally we  present our leader selection approach and derive a bound on the regret achieved by our approach.

\subsubsection{Problem Formulation}
 If a fixed set of leaders is maintained for the lifetime of the network, then high convergence errors will result since the future topologies are not known in advance.  Hence,
 we assume that a new leader set $S_{m}$ is selected for each time interval $[t_{m}, t_{m+1}]$.  We note that the node dynamics defined in Section \ref{subsec:model} will converge to $\mathbf{x}^{\ast}$ if $\mathbf{x}^{\ast} = x^{\ast}\mathbf{1}$, i.e., if all leader agents maintain the same state $x^{\ast}$.    For this case, letting $\delta_{m} = t_{m} - t_{m-1}$, and recalling that $r = \max{\{m: t_{m} \leq t\}}$ represents the number of topology changes before time $t$,
we have the following upper bound on the convergence error for dynamic networks.
\begin{proposition}
\label{prop:unknown_convergence_error}
For any topology dynamics $G(t)$,
\begin{multline}
\label{eq:error_metric_upper_bound}
\sum_{i=1}^{n}{\left[\sum_{j \in V \setminus S}{\left(e_{i}^{T}\prod_{m=1}^{r}{e^{-L(t_{m-1})\delta_{m}}}\right)_{j}^{p}} + \left(1 - \sum_{j \in S}{\left(e_{i}^{T}\prod_{m=1}^{r}{e^{-L(t_{m-1})\delta_{m}}}\right)_{j}}\right)^{p}\right]} \\
 \leq \sum_{m=1}^{r}{\sum_{i \in V \setminus S_{m}}{\left[\sum_{j \in V \setminus S}{\left(e_{i}^{T}e^{-L(t_{m-1})\delta_{m}}\right)_{j}^{p}} + \left(1 - \sum_{j \in S}{\left(e_{i}^{T}e^{-L(t_{m-1})\delta_{m}}\right)_{j}}\right)^{p}\right]}}.
\end{multline}
\end{proposition}
\begin{IEEEproof}
  Since $e_{0}^{T}$ is an absorbing state of the Markov chain with transition matrix $e^{-L_{m}(S_{m})\delta_{m}}$ for all $m \in \{1, \ldots, r\}$, $e_{0}^{T}e^{-L_{m}(S_{m})\delta_{m}} = e_{0}^{T}$ for all $m \in \{1, \ldots, r\}$.  For any $i \in V$, this implies
  \begin{IEEEeqnarray}{rCl}
  \nonumber
  ||e_{i}^{T}\prod_{m=1}^{r}{e^{-L_{m}(S_{m})\delta_{m}}} - e_{0}^{T}||_{p}^{p} &=& ||(e_{i}e^{-L_{r}(S_{r})\delta_{r}} - e_{0}^{T})e^{-L_{r-1}(S_{r-1})\delta_{r-1}}\cdots e^{-L_{1}(S_{1})\delta_{1}}||_{p}^{p} \\
  \label{eq:unknown_helper_bound1}
  &\leq& ||e_{i}^{T}e^{-L_{r}(S_{r})\delta_{r}} - e_{0}^{T}||_{p}^{p}||e^{-L_{r-1}(S_{r-1})\delta_{r-1}}||_{p}^{p}\cdots ||e^{-L_{1}(S_{1})\delta_{1}}||_{p}^{p} \\
  \label{eq:unknown_helper_bound2}
  &\leq& ||e_{i}^{T}e^{-L_{r}(S_{r})\delta_{r}} - e_{0}^{T}||_{p}^{p},
  \end{IEEEeqnarray}
  where (\ref{eq:unknown_helper_bound1}) follows from the definition of the matrix $2$-norm, while (\ref{eq:unknown_helper_bound2}) follows from the fact that $e^{-L_{m}(S_{m})\delta_{m}}$ is a stochastic matrix for all $m \in \{1, \ldots, r-1\}$.  Eq. (\ref{eq:unknown_helper_bound2}), together with the fact that $||e_{i}^{T}e^{-L_{m}(S_{m})\delta_{m}} - e_{0}^{T}||_{p}^{p} \geq 0$ for all $m \in \{1, \ldots, r-1\}$, yields the desired result.
\end{IEEEproof}

Based on Proposition \ref{prop:unknown_convergence_error}, we define an error metric for the unknown topology case by
\begin{equation}
\label{eq:dynamic_metric}
\hat{f}_{t}(S_{1},\ldots,S_{r}) \triangleq \frac{1}{r}\sum_{m=1}^{r}{\sum_{i \in V \setminus S_{m}}{\left[\sum_{j \in V \setminus S}{\left(e_{i}^{T}e^{-L(t_{m-1})\delta_{m}}\right)_{j}^{p}} + \left(1 - \sum_{j \in S}{\left(e_{i}^{T}e^{-L(t_{m-1})\delta_{m}}\right)_{j}}\right)^{p}\right]}}.
\end{equation}
Using the metric (\ref{eq:dynamic_metric}), minimizing the convergence error is achieved by selecting leaders according to the optimization problem
\begin{equation}
\label{eq:unknown_dynamics_opt}
\begin{array}{ll}
\mbox{minimize} \vspace{-0.1in} & \hat{f}_{t}(S_{1},\ldots,S_{r}) \\
S_{1}, \ldots, S_{r} \vspace{-0.1in} & \\
\mbox{s.t.} & |S_{m}| \leq k, \ m=1,\ldots,r
\end{array}
\end{equation}
At time $t_{m}$, the system designer has access to the sequence of Laplacian matrices $L(t_{1}), \ldots, L(t_{m-1})$, and hence can compute the convergence error bound $\hat{f}_{t}(S_{1},\ldots,S_{m-1})$ arising from any sequence of leader sets $S_{1}, \ldots, S_{m-1}$.

In order to analyze the optimality of our proposed algorithm and prove lower bounds on the achievable optimality gap, we introduce the concept of regret, denoted $R(S_{1},\ldots,S_{r})$, which is defined as the difference between the convergence error from sets $S_{1},\ldots,S_{r}$ and the minimum convergence error from any fixed leader set.  The regret is defined as
\begin{equation}
\label{eq:regret_def}
R(S_{1}, \ldots, S_{r}) =   \hat{f}_{t}(S_{1}, \ldots, S_{r}) - \min_{S}{\left\{\frac{1}{r}\sum_{m=1}^{r}{\hat{f}_{t}(S|L(t_{m}))}\right\}}.
\end{equation}
The lower bounds and optimality gap derived below are based on the regret (\ref{eq:regret_def}).

\subsubsection{Lower bounds on regret}
In what follows, we give a lower bound on the minimum possible regret for any algorithm without knowledge of the topology distribution.  This bound provides a comparison for evaluating possible leader selection algorithms, including our proposed approach.  In order to prove the bound, we construct a lower bound on the regret for a specific topology distribution in which, at each time epoch, each node is independently connected to all other nodes with probability $1/2$, and only one node can act as leader.  Intuitively, the leader selected by any algorithm without topology information will provide very high convergence error with probability $1/2$, leading to a large regret.  On the other hand, an algorithm with knowledge of the network topology will choose a leader node that is connected to all other nodes, with low resulting convergence error.  Formally, the theorem is stated as follows.

\begin{theorem}
\label{theorem:regret_lower_bound}
For any leader selection algorithm for solving (\ref{eq:unknown_dynamics_opt}) with $k=1$ and for $n$ and $r$ sufficiently large, there exists a sequence of topologies $G(t_{1}), \ldots, G(t_{r})$ such that the regret is bounded by
\begin{equation}
\label{eq:regret_lower_bound}
R \geq \frac{1}{r}\sqrt{r/2 \ln{n}}.
\end{equation}
\end{theorem}
A proof is given in the appendix.

\subsubsection{Experts Algorithms}
We now give background on experts algorithms, which will be used as subroutines in our algorithms.  All results can be found in \cite{cesa2006prediction}.  As a preliminary, we define a set of $n$ actions $\mathcal{A} = \{a_{1}, \ldots, a_{n}\}$ (we use the notation $n$ because, in the next subsection, each expert will correspond to selecting a different node as leader).  We also define a sequence of loss functions $\ell_{1},\ldots,\ell_{T}: \mathcal{A} \rightarrow \mathbb{R}_{+}$ for time epochs $1,\ldots,T$.  The loss for action $a_{j}$ in epoch $i$ is defined to be $l_{i}(a_{j})$.

An \emph{experts algorithm} outputs an action $a \in \mathcal{A}$ at each epoch $i$ based on observations of past losses $\{l_{i^{\prime}}(a) : a \in \mathcal{A}, i^{\prime} = 1,\ldots,i-1\}$, in order to minimize the total losses.  The effectiveness of an experts algorithm can be quantified using the regret, defined as follows.
\begin{definition}
\label{def:regret_general}
The regret of an experts algorithm that chooses action $a^{(j)}$ at time epoch $j$ is defined as $$R \triangleq \sum_{i=1}^{T}{\ell_{i}(a^{(j)})} - \min_{a \in \mathcal{A}}{\left\{\sum_{i=1}^{T}{\ell_{i}(a)}\right\}}.$$
\end{definition}
One such experts algorithm is given below as Algorithm \emph{Randomized-experts}.

\begin{figure}[h]
\centering
\renewcommand{\arraystretch}{0.6}
\begin{tabular}{l}
\hline
\small{\textbf{Randomized-experts}: Experts algorithm with sublinear regret.}\\
\small{\textbf{Input:} Set of actions $\mathcal{A} = \{a_{1},\ldots,a_{n}\}$}\\
\small{\textbf{Output:} Actions $a^{(1)}, \ldots, a^{(T)}$ for epochs $1,\ldots,T$}\\
\small{\textbf{Initialize:} Parameter $\eta \in [0,1]$, $\mathbf{w}_{j} \leftarrow 1$, $j=1,\ldots,n$, $\mathbf{p}_{j} \leftarrow \frac{1}{n}$, $j=1,\ldots,n$}\\
\small{\textbf{for each epoch} $i=1,\ldots,T$}\\
\small{~~Select $a^{(i)}$ from probability distribution $\mathbf{p}$}\\
\small{~~Receive losses $\ell_{i}(a_{1}),\ldots,\ell_{i}(a_{n})$}\\
\small{~~\textbf{for} $j=1,\ldots,n$} \\
\small{~~~~$\mathbf{w}_{j} \leftarrow \mathbf{w}_{j}\exp{(-\eta \ell_{i}(a_{j}))}$} \\
\small{~~\textbf{end for}}\\
\small{~~$\mathbf{p} \leftarrow \left(\mathbf{1}^{T}\mathbf{w}\right)^{-1}\mathbf{w}$}\\
\small{\textbf{end for}}\\
\hline
\end{tabular}
\end{figure}

Intuitively, the \emph{Randomized-experts} algorithm maintains a probability distribution $\mathbf{p} \in \mathbf{R}^{n}$, where $\mathbf{p}_{j}$ represents the probability of selecting action $j$.  The algorithm assigns higher weight, and hence higher selection probability, to actions that have generated low losses in the past.  \textbf{Randomized-experts} updates the weights according to an exponential rule.  While alternative update algorithms such as polynomial weighting have been studied, exponential weighting schemes have been found to provide lower error in empirical and analytical studies~\cite{cesa2006prediction}.
The following proposition characterizes the regret of \emph{Randomized-experts}.
\begin{proposition}
\label{prop:regret_experts}
The regret of the set $a^{(1)},\ldots, a^{(T)}$ returned by the algorithm \emph{Randomized-experts} satisfies $R(a^{(1)},\ldots,a^{(T)}) \leq O\left(\sqrt{\frac{T \ln{n}}{2}}\right)$.
\end{proposition}

\subsubsection{Leader selection algorithms for unknown topology distribution} Our approach to solving the optimization problem (\ref{eq:unknown_dynamics_opt}) is based on the greedy algorithm \emph{Select-$k$-leaders}.  In the unknown topology setting, however, the exact value of $\hat{f}_{t}(S)$ cannot be computed.  Instead, we use the experts algorithm \emph{Randomized-experts} to estimate which leader $v_{j}$ to add to the leader set at the $j$-th iteration of the algorithm.

  The algorithm maintains a set of weights $w_{ij}^{m}$ for $i = 1, \ldots, n$, $j = 1, \ldots, k$, and each time step $m=1,\ldots,r$.  The weight $w_{ij}^{m}$ represents the current estimate of node $i$'s utility when selected as the $j$-th leader node at time $t_{m}$.    Initially, $w_{ij}^{0} = 1$ for all $i$ and $j$, since no information is available regarding the effectiveness of each node as a leader.  At step $m$, a set of probability distributions $\pi_{1}^{m}, \ldots, \pi_{k}^{m}$ is obtained by setting $\pi_{j}^{m}(i) = w_{ij}^{m}/\sum_{i=1}^{n}{w_{ij}^{m}}$.  The leader set $S_{m}$ for time interval $[t_{m},t_{m+1}]$ is obtained by first selecting a leader according to distribution $\pi_{1}^{m}$.  In general, the $j$-th leader is selected by choosing a node from $V \setminus S_{j-1}$ according to distribution $\pi_{j}^{m}$, which can be viewed as selecting the leader according to an experts algorithm where the weights are computed based on the convergence error during previous time steps.  

  After stage $m$, the weights are updated in order to reflect the convergence error that each node would have provided if it had acted as leader during $[t_{m},t_{m+1}]$.   Define $z_{m,i,j}$ to be $\hat{f}_{t}(S_{m}^{j-1}) - \hat{f}_{t}(S_{m}^{j-1} \cup \{i\})$.  The weight $w_{ij}^{m}$ is updated to $w_{ij}^{m+1} = \beta^{z_{m,i,j}}w_{ij}^{m}$, as in algorithm \textbf{Randomized-experts}, where $\beta \in [0,1]$ is chosen in order to vary the rate at which the algorithm adapts to changes in topology.  For low $\beta$ values, a large convergence error will result in a large penalty (i.e., a much lower weight) for a node.  The algorithm is summarized as \emph{Select-dynamic-leaders} below.

  \begin{figure}[h]
  \centering
  \renewcommand{\arraystretch}{0.6}
  \begin{tabular}{l}
  \hline
  \small{\textbf{Select-dynamic-leaders}: Algorithm for selecting up to}   \small{$k$ leaders for time interval $[t_{r}, t_{r+1}]$ when the} \\
  \small{topology dynamics are unknown.}  \\
  \small{\textbf{Input:} Set of nodes $V = \{1, \ldots, n\}$},  \small{maximum number of leaders $k$} \\
  \small{Node weights $w_{ij}^{r-1}$, $i \in V$, $j=1,\ldots,k$},  \small{topology $G(t_{r})$}\\
  \small{\textbf{Output:} Leader set $S_{r}$ for state $r$}, \small{updated weights $w_{ij}^{r}$, $i \in V$, $j=1,\ldots,k$}\\
  \small{\textbf{Initialization:} $z_{ij}^{r} \leftarrow 0$, $i \in V$, $j=1,\ldots,k$}, \small{$S_{r} \leftarrow \emptyset$} \\
  \small{\textbf{for}($j=1:k$)}\\
  \small{~~~\textbf{for}($i=1:n$)}\\
  \small{~~~~~~$z_{ij}^{r} \leftarrow \hat{f}_{t}(S_{r-1}^{j-1}) - \hat{f}_{t}(S_{r-1}^{j-1} \cup \{i\})$}, \small{~~~~$w_{ij}^{r} \leftarrow w_{ij}^{r-1}\beta^{z_{ij}^{r}}$} \quad //Based on experts algorithm\\
  \small{~~~\textbf{for}($i=1:n$)}\\
  \small{~~~~~~$\pi_{ij}^{r} \leftarrow w_{ij}^{r}/\left(\sum_{v \in V}{w_{vj}^{r}}\right)$}\\
  \small{~~~$v_{j}^{r} \leftarrow \mbox{choose } i \in V \setminus S_{r} \mbox{ according to } \pi_{j}^{r}$}\\
  \small{~~~$S_{r} \leftarrow S_{r} \cup \{v_{j}^{r}\}$}\\
  \small{\textbf{end for}}\\
  \small{\textbf{return} $S_{r}$}\\
  \hline
  \end{tabular}
  \end{figure}

  The following theorem describes bounds achieved on the regret of the algorithm \emph{Select-dynamic-leaders}.
  \begin{theorem}
  \label{theorem:unknown_dynamics_error}
  The algorithm \emph{Select-dynamic-leaders} returns a sequence of sets $S_{1},\ldots,S_{r}$ that satisfies
  \begin{displaymath}
  \label{eq:unknown_dynamics_error}
  \hat{f}_{t}(S_{1},\ldots,S_{r}) \leq \left(1 - \frac{1}{e}\right)\hat{f}_{t}(S^{\ast}) + \frac{1}{e}f_{max} + \sum_{j=1}^{k}{R_{j}},
  \end{displaymath}
  where $S^{\ast} = \arg\min{\left\{\sum_{m=1}^{r}{\hat{f}_{t}(S|L(t_{m}))} : |S| \leq k\right\}}$ and the regret $R_{j}$ in choosing the $j$-th leader is given by
  $R_{j} \leq \frac{1}{r}\left(\sqrt{2f_{max}r\ln{n}} + \ln{n}\right)$.
  \end{theorem}

  \begin{IEEEproof}
  By Theorem 6 of \cite{streeter2007online}, an algorithm for maximizing a submodular function $f(S)$ that introduces error $R_{j}$ from the greedy approach at each stage $j$, i.e.,
  \begin{displaymath}
  \max_{v}{f(S \cup \{v\})} - f(S \cup \{v_{j}\}) \leq R_{j}
  \end{displaymath}
  satisfies
  \begin{displaymath}
  f(S) > \left(1 - \frac{1}{e}\right)f(S^{\ast}) - \sum_{j=1}^{k}{R_{j}}.
  \end{displaymath}
  Using the fact that $\hat{f}_{t}(S_{r})$ is supermodular as a function of $S_{r}$ (Theorem \ref{theorem:main_result}), setting $f(S) = f_{max} - \hat{f}_{t}(S)$ and rearranging terms implies that
  \begin{displaymath}
  \hat{f}_{t}(S_{r}) \leq \left(1 - \frac{1}{e}\right)\hat{f}_{t}(S_{r}^{\ast}) + \frac{1}{e}f_{max} + \sum_{j=1}^{k}{R_{j}}.
  \end{displaymath}
  It remains to bound the regret $R_{j}$.  Lemma 4 of \cite{freund1995desicion} yields the desired bound.
  \end{IEEEproof}

Theorem \ref{theorem:unknown_dynamics_error} implies that, as the number of topologies $r$ increases, the convergence error of algorithm \emph{Select-dynamic-leaders} reaches a constant factor of the optimal convergence error.  Indeed, \emph{Select-dynamic-leaders} achieves
the same optimality gap of $\left(1 - \frac{1}{e}\right)$ as the known topology distribution case.  In other words, if the system designer observes the distribution of the network topology for a sufficiently long time, then the convergence error with the chosen leader set is equivalent to the convergence error when the topology distribution is known in advance. The algorithm incurs a cost of $O(nk)$ computations of $\hat{f}_{t}(S)$ for each network topology.


%% file: Simulation.tex
\section{Numerical Study}
\label{sec:simulation}
We conduct a numerical study of our proposed approach using Matlab.  We simulate a network of $100$ nodes, deployed uniformly at random over a $1000$m $\times$ $1000$m area.  A communication link exists between two nodes if they are within $300$m of each other.  The weight matrix $W$ of Equation (\ref{eq:follower_dynamics}) is chosen by selecting a weight $W_{ij}$ for each link $(i,j)$ uniformly from the interval $[0,50]$.  Each data point represents an average of $50$ independent trials.  The number of leaders varies from $1$ to $15$.


Three types of network topologies are considered.  The first type consists of a static network topology with the parameters described above.  In the second type, a topology with the parameters described above is subject to link failures.  As a result, the topology during time interval $[t_{m},t_{m+1}]$, with $m=1,\ldots,8$,  is obtained by removing each link independently and with equal probability from the underlying network.  The probability of link failures varies from $0$ to $0.15$.  In the third type of network topology, the positions of the network nodes vary according to a random waypoint mobility model~\cite{hong1999group}.  In this model, the nodes attempt to maintain a fixed position relative to a time-varying reference state.  During each time interval, the position of each node is equal to the sum of the reference state, the desired relative position of that node, and a disturbance with magnitude chosen uniformly at random from the interval $[0,50]$.  The relative positions are chosen uniformly at random from the $1000$m $\times$ $1000$m area, while the reference state moves according to a random walk with speed $100$m/s.  The number of time epochs varies from $1$ to $8$ in order to track the performance of each simulated algorithm over time.

For each type of topology, leader selection is performed using four algorithms, namely, random selection of leader nodes, selection of maximum-degree nodes as leaders, selection of average-degree nodes, and leader selection via supermodular optimization.  For the problem of selecting up to $k$ leaders in order to minimize convergence errors, the algorithm \emph{Select-$k$-leaders} is used, while algorithm \emph{Select-minimal-leaders} is used to select the minimum-size set of leaders in order to achieve a convergence error bound.  In addition, for the two types of dynamic topologies that are considered, a comparison between the algorithm \emph{Select-$k$-leaders}, which incorporates prior knowledge of the topology distribution, and the algorithm \emph{Select-dynamic-leaders}, which does not incorporate prior knowledge, is provided.


The goal of the numerical evaluation is to address the following questions: (i) How does the convergence error of the leaders selected using supermodular optimization compare with state of the art leader selection algorithms? (ii) How does the number of leaders required by each scheme vary as a function of the convergence error bound?  and (iii) When the topology distribution is unknown, does the algorithm \emph{Select-dynamic-leaders} learn the distribution and achieve convergence error comparable to the known topology case?

\begin{figure}
\centering
$\begin{array}{cc}
\includegraphics[width=3in]{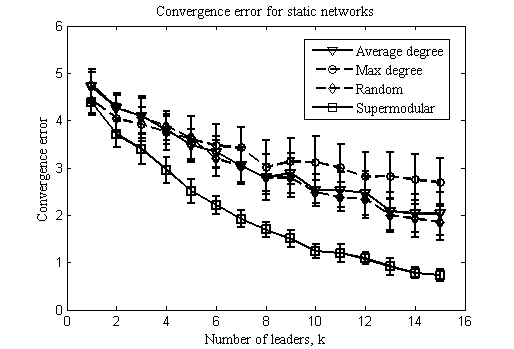} &
\includegraphics[width=3in]{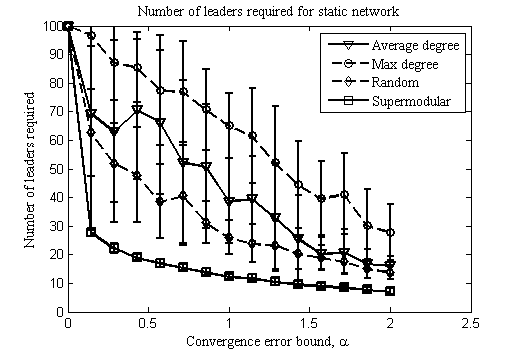} \\
(a) & (b)
\end{array}$
\caption{Comparison of random, maximum degree, average degree, and supermodular leader selection algorithm \emph{Select-$k$-leaders} for static networks. (a)  The supermodular optimization approach \emph{Select-$k$-leaders} consistently provides the lowest convergence error, while the average degree-based selection slightly outperforms random leader selection. (b) Evaluation of the selection of the minimum-size set of leader nodes to achieve a given bound on the convergence error, as a function of the bound, for static networks.  The supermodular optimization approach \emph{Select-minimal-leaders} requires fewer leaders in order to satisfy convergence error criteria.}
\label{fig:simulation_static}
\end{figure}

\begin{figure*}
\centering
$\begin{array}{cc}
\includegraphics[width=3in]{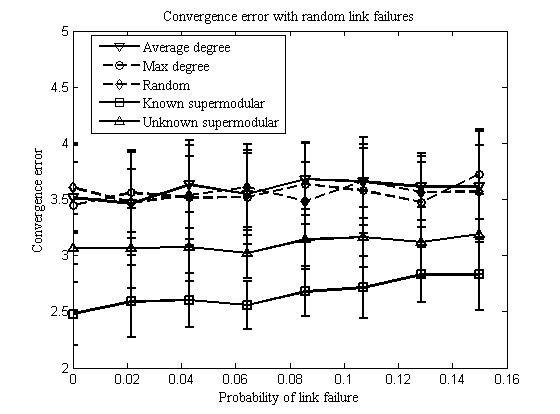} &
\includegraphics[width=3in]{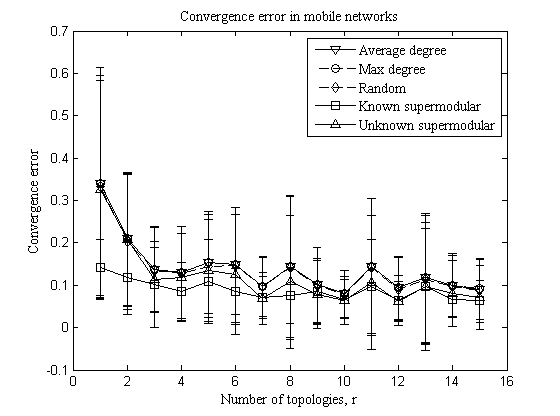} \\
(a) & (b)
\end{array}$
\caption{Comparison of random, degree-based, and supermodular optimization algorithms \emph{Select-dynamic-leaders} and \emph{Select-$k$-leaders} for dynamic networks.  (a) Impact of random link failures on convergence errors.  The convergence error for each scheme increases along with the failure probability.  Furthermore, for each failure probability, knowledge of the distribution of link failures leads to lower convergence error than the unknown topology case.  (b) Convergence errors when the network topology varies due to random waypoint mobility model with speed $v=100$m/s.  As the number of topologies increases, the algorithm \emph{Select-dynamic-leaders} adaptively updates the leader set, eventually providing error comparable to the known topology case.}
\label{fig:simulation_dynamic}
\end{figure*}

Question (i) is addressed for the static network case in Figure \ref{fig:simulation_static}(a).  The supermodular optimization approach of algorithm \emph{Select-$k$-leaders} selects leaders with less than half the convergence error of random leader selection.  We observe that, while the random, degree, and average degree-based algorithms achieve comparable convergence error, selection of random nodes as leaders slightly outperforms both degree-based schemes.  Similarly, in Figures \ref{fig:simulation_dynamic}(a) and \ref{fig:simulation_dynamic}(b) supermodular optimization of the leader set, using the algorithms \emph{Select-$k$-leaders} and \emph{Select-dynamic-leaders}, results in lower convergence error than random and degree-based schemes for dynamic topologies.

Figure \ref{fig:simulation_static}(b) is related to Question (ii).  The supermodular optimization approach of algorithm \emph{Select-minimal-leaders} requires less than half the number of leaders to achieve an error bound of $1$ than the next highest-performing scheme, which was random leader selection.  As in the case of selecting a fixed number of leaders, random leader selection outperformed degree-based leader selection on average.

Figures \ref{fig:simulation_dynamic}(a) and \ref{fig:simulation_dynamic}(b) compare the convergence error for the supermodular optimization algorithms, with and without prior topology information (Question (iii)).  In the case of random link failures (Figure \ref{fig:simulation_dynamic}(a)), prior knowledge in algorithm \emph{Select-$k$-leaders} provides a consistent advantage over the case where no prior information is available, using \emph{Select-dynamic-leaders}.  This is because the link failures occur independently at each time interval, and hence the ability of the algorithm \emph{Select-dynamic-leaders} to dynamically adapt to new topology information does not lead to lower convergence error.  When the topology changes due to a mobility model (Figure \ref{fig:simulation_dynamic}(b)), the \emph{Select-dynamic-leaders} algorithm eventually achieves the same performance as the algorithm with prior knowledge as time progresses.  Given sufficient time intervals, \emph{Select-dynamic-leaders} eventually learns which leader nodes will provide the lowest convergence error under the  mobility model that is used.

%% file: Conclusion.tex
\section{Conclusion}
\label{sec:conclusion}
In this paper, we investigated leader selection for minimizing convergence error, defined as the distance between the intermediate states of the follower agents and the convex hull of the leader states, in linear multi-agent systems.  We developed efficient algorithms for leader selection through the following approach.  First, we derived an upper bound on the convergence error  that is independent of the initial states of the network, and proved a connection between the upper bound and a random walk on the graph.   Using the connection between convergence error and the random walk,  we proved that the upper bound on the convergence error is a supermodular function of the set of leader agents. The supermodular structure of the convergence error enables formulation and approximation of leader selection as a discrete optimization problem, rather than relying on continuous extensions of the problem.  

We formulated two leader selection problems for MAS with static topology.  In the first problem, a fixed number of leaders is selected in order to minimize the convergence error.  In the second problem, the minimum-size set of leaders is selected in order to achieve a given bound on the convergence error.  We presented efficient algorithms for each problem, and proved that both algorithms achieve an $O(1)$ optimality gap with the lowest possible convergence error.

We introduced a supermodular optimization approach to leader selection in MAS with dynamic topologies, including the cases where the system designer has prior knowledge of the distribution of the topology, as well as the case where the system designer has no prior information and must adaptively update the leader set over time.  For the case where the system designer has no prior topology information, we derived a lower bound on the convergence error that can be achieved by any algorithm, as well as bounds on the convergence error achieved by our approach.

  Our results were illustrated through a numerical study, in which we  compared our supermodular optimization approach with random, average-degree, and maximum-degree leader selection, for static networks, networks with random link failures, and networks with mobile agents.  We found that the supermodular optimization approach significantly outperforms the other algorithms.  Furthermore, the numerical evaluation showed that the  leader selection approach without prior topology information eventually achieves a convergence error that is comparable to the algorithm with prior information.  

  In our future work, we will investigate distributed approaches to leader selection in order to minimize convergence error.  Such a distributed approach must take into account the computation, bandwidth, and energy constraints of the distributed agents, as well as the limited, local information available to each agent.  Our approach will attempt to leverage the supermodular structure of the convergence error in order to  develop polynomial-time approximation algorithms that satisfy these constraints. 

%% file: Proofs.tex
\section*{Appendix}
\label{sec:proofs}

In this appendix, proofs of Lemma \ref{lemma:monotone}, Theorem \ref{theorem:supermodular_dynamic},  and Theorem \ref{theorem:regret_lower_bound} are given.

\begin{IEEEproof}[Proof of Lemma \ref{lemma:monotone}]
We have that $\mathbf{x}(t) = e^{-L(t-t_{0})}\mathbf{x}(t_{0})$.  Since $e^{-L(t-t_{0})}$ is a stochastic matrix, we have that, for all $i \in V \setminus S$, $x_{i}(t) = \sum_{j \in V}{\alpha_{ij}x_{j}(t_{0})}$ for some $\alpha_{ij} \geq 0$ satisfying $\sum_{j \in V}{\alpha_{ij}} = 1$ (also, by definition of $S$, we have that $\alpha_{ij} \equiv 0$ for all $i \in S$, $j \neq i$).  Let $y_{j}^{0} \in \arg\min{\{|x_{j}(t_{0}) - y|^{p} : y \in \overline{A}\}}$, and define $\overline{y}_{i} = \sum_{j \in V}{\alpha_{ij}y_{j}^{0}}$, noting that $\overline{y}_{i} \in \overline{A}$.  We have
\begin{IEEEeqnarray*}{rCl}
|x_{i}(t) - \overline{y}_{i}|^{p} &=& \left|\sum_{j \in V}{\alpha_{ij}x_{j}(t_{0})} - \sum_{j \in V}{\alpha_{ij}y_{j}^{0}}\right|^{p} = \left|\sum_{j \in V}{\alpha_{ij}(x_{j}(t_{0}) - y_{j}^{0})}\right|^{p} \\
&\leq& \sum_{j \in V}{\alpha_{ij}|x_{j}(t_{0}) - y_{j}^{0}|^{p}} \leq \sum_{j \in V \setminus S}{\alpha_{ij}|x_{j}(t_{0}) - y_{j}^{0}|^{p}},
\end{IEEEeqnarray*}
where the first inequality follows from convexity of $|\cdot|^{p}$.  We then have
\begin{IEEEeqnarray*}{rCl}
f_{t}(S) &=& \left(\sum_{i \in V}{\min_{y \in \overline{A}}{\{|x_{i}(t) - y|^{p}\}}}\right)^{1/p} \leq \left(\sum_{i \in V}{|x_{i}(t) - \overline{y}_{i}|^{p}}\right)^{1/p} \\
&\leq& \left(\sum_{i \in V \setminus S}{\sum_{j \in V \setminus S}{\alpha_{ij}|x_{j}(t_{0})-y_{j}^{0}|^{p}}}\right)^{1/p} = \left(\sum_{i \in V \setminus S}{\sum_{j \in V \setminus S}{\alpha_{ji}|x_{j}(t_{0}) - y_{j}^{0}|^{p}}}\right)^{1/p} \\
&=& \left(\sum_{j \in V \setminus S}{\left(|x_{j}(t_{0}) - y_{j}^{0}|^{p}\sum_{i \in V \setminus S}{\alpha_{ji}}\right)}\right)^{1/p} \leq \left(\sum_{j \in V \setminus S}{\left(|x_{j}(t_{0}) - y_{j}^{0}|^{p}\right)}\right)^{1/p} = f_{t_{0}}(S),
\end{IEEEeqnarray*}
as desired.
\end{IEEEproof}

  \begin{IEEEproof}[Proof of Theorem \ref{theorem:supermodular_dynamic}]
Suppose that the topology changes occur at given times $t_{1} <  \ldots < t_{r}$, with $\delta_{m} = t_{m}-t_{m-1}$.  Choose $\delta$ sufficiently small, and let $\tau_{0} = \lceil \frac{t_{1}-t_{0}}{\delta}\rceil$, $\ldots$, $\tau_{r} = \lceil \frac{t-t_{r}}{\delta}\rceil$.  Note that the difference between $\tau_{m}\delta$ and $(t_{m+1}-t_{m})$ can be made arbitrarily small by decreasing $\delta$.  Since $P_{\delta}^{m}$ (where $P_{\delta} = e^{-L\delta}$) is a stochastic matrix for each $m$, the product $\prod_{m=1}^{r}{P_{\delta}^{m}}$ is also stochastic.
Let $X(l)$ represent a random walk on $V$ such that the $(\tau_{m-1})$-th to $(\tau_{m})$-th steps are taken  using transition matrix $P_{\delta}^{m}$.

We first show that the term of the inner summation of (\ref{eq:hat_epsilon_dynamic}) with indices $i$ and $j$ is equivalent to the probability that $X(t_{r})$ is equal to $j$ when the random walk starts at $i$.  Formally, in the limit as $\delta \rightarrow 0$, this relation is given by
\begin{equation}
\label{eq:dynamic_first_step}
\left(e_{i}^{T}\prod_{m=1}^{r}{e^{-L(t_{m-1})\delta_{m}}}\right)_{j} = Pr(X(\tau_{0}+ \tau_{1} + \cdots + \tau_{r}) = j | X(0) = i).
\end{equation}
The proof of (\ref{eq:dynamic_first_step}) is given by induction on $r$.  When $r=1$, the proof follows from Theorem \ref{theorem:random_walk}.  Now, assume that (\ref{eq:dynamic_first_step}) holds for $r < r_{0} \in \mathbb{Z}_{+}$, and define the probability distribution $\pi$ by $\pi^{T} = e_{i}^{T}\prod_{m=1}^{r_{0}-1}{e^{-L(t_{m-1})\delta_{m}}}$.  This definition yields
\begin{IEEEeqnarray}{rCl}
\nonumber
\left(e_{i}^{T}\prod_{m=1}^{r_{0}}{e^{-L(t_{m-1})\delta_{m}}}\right)_{j} &=& \sum_{j^{\prime}=0}^{n}{\pi_{j^{\prime}}(e^{-L(t_{r_{0}-1})(t_{r_{0}}-t_{r_{0}-1})})_{j^{\prime}j}} \\
\nonumber
&=& \sum_{j^{\prime}=0}^{n}{\left[Pr(X(\tau_{r_{0}-1}+ \cdots + \tau_{0}) = j^{\prime} | X(0) = i) \times \right.} \\
\label{eq:supermodular_dynamic_helper}
 && \left.Pr(X(\tau_{r_{0}}) = j | X(0) = j^{\prime})\right] \\
\nonumber
&=& \sum_{j^{\prime}=0}^{n}{\left[Pr(X(\tau_{r_{0}-1} + \cdots + \tau_{0}) = j^{\prime} | X(0) = i) \times \right. } \\
\label{eq:supermodular_dynamic_helper1}
 && \left. Pr(X(\tau_{0} + \cdots + \tau_{r_{0}}) = j | X(\tau_{0} + \cdots + \tau_{r_{0}-1}) = j^{\prime})\right],
\end{IEEEeqnarray}
where (\ref{eq:supermodular_dynamic_helper}) follows by using the inductive assumption and Theorem \ref{theorem:random_walk}.  Equation (\ref{eq:supermodular_dynamic_helper1}) follows from the stationarity of the random walk. 
 Equation (\ref{eq:supermodular_dynamic_helper1}) and the law of total probability then imply (\ref{eq:dynamic_first_step}).  Hence $||e_{i}^{T}\prod_{m=1}^{r}{e^{-L(t_{m-1})\delta_{m}}}||_{p}^{p}$ is supermodular by (\ref{eq:dynamic_first_step}) and Lemma \ref{lemma:main_result_helper}.  The function $\hat{f}_{t}(S)$ is therefore a nonnegative weighted sum of supermodular functions, and hence is supermodular.
\end{IEEEproof}


   \begin{IEEEproof}[Proof of Theorem \ref{theorem:regret_lower_bound}]
We prove the theorem by constructing a sequence of random topologies $(G(t_{1}), \ldots,$ $G(t_{r}))$ such that, when $k=1$, the expected regret of any leader selection algorithm that does not take the distribution of the network topology as input satisfies (\ref{eq:regret_lower_bound}).  An outline of the proof is as follows.  We begin by defining the distribution of $G(t_{1}), \ldots, G(t_{r})$. The next step is to derive an expression for $R$ for this choice of topologies.  We then prove the bound (\ref{eq:regret_lower_bound}) by analyzing the asymptotic behavior of our expression for $R$.  Details of the proof are given below.

 Consider a directed graph $G(t)$ in which, at time  $t_{m}$, the edge set $E(t_{m})$ is chosen such that, for each node $i$, the neighbor set $N(i)$ satisfies
\begin{displaymath}
N(i) = \left\{
\begin{array}{cc}
V \setminus \{i\}, & \mbox{w.p. } 1/2 \\
\emptyset, & \mbox{w.p. } 1/2
\end{array}
\right.
\end{displaymath}
 In other words, each node has outdegree $(n-1)$ with probability $1/2$ and outdegree $0$ with probability $\frac{1}{2}$.  Define $\sigma(n)$ to be the normalized expected convergence error when a node with outdegree $(n-1)$ acts as leader.  The value of $\sigma(n)$ is normalized so that the convergence error is $1$ when the outdegree of the leader is $0$.  This yields 
   \begin{equation}
    \label{eq:minimize_one_leader}
   \hat{f}_{t}(\{i\}|G(t_{m})) = \left\{
   \begin{array}{cc}
   \sigma(n), & \mbox{w.p. } 1/2 \\
   1, & \mbox{w.p. } 1/2.
   \end{array}
   \right.
   \end{equation}
   It is assumed that each node's edge set is chosen independently at random at each time step.

   The first term of $R$ in (\ref{eq:regret_def}) is given as follows.
   Any algorithm that does not have foreknowledge of the network topology will have expected error that satisfies 
\begin{IEEEeqnarray*}{rCl}
\nonumber
\mathbf{E}(\hat{f}_{t}(S_{1}, \ldots, S_{r})) &=& \sum_{m=1}^{r}{\sum_{i=1}^{n}{\mathbf{E}(\hat{f}_{t}(\{i\})|S_{m} = \{i\})Pr(S_{m} = \{i\})}} \\
\label{eq:expected_error}
&=& \sum_{m=1}^{r}{\left[\frac{1}{2}(1+\sigma(n))\sum_{i=1}^{n}{Pr(S_{m} = \{i\})}\right]} = \frac{r}{2}(1 + \sigma(n)).
\end{IEEEeqnarray*}

 We now consider the second term of $R$ in (\ref{eq:regret_def}). Let $A_{r,i} \triangleq \sum_{m=1}^{r}{\hat{f}_{t}(\{i\}|G(t_{m}))}$ to be the convergence error for $r$ topologies when the leader node is $i$, so that the second term of (\ref{eq:regret_def}) is equal to $\min_{i}{A_{r,i}}$.   The mean of $A_{r,i}$ is $\frac{r}{2}(1 + \sigma(n))$ from (\ref{eq:minimize_one_leader}), and the variance is $\frac{r}{4}(1-\sigma(n))^{2}$.  When $\sigma(n) = 0$, the value of $A_{r,i}$ is minimized, and hence the second term of (\ref{eq:regret_def}) is minimized.  In what follows, we therefore assume that $\sigma(n) = 0$.  
 Under this assumption, $A_{r,i}$ is a binomial random variable.   
We now have that the expected value of the regret, when $\sigma(n) = 0$, is equivalent to $R = \frac{1}{r}(\frac{r}{2} - \min_{i}{A_{r,i}})$, so that (\ref{eq:regret_lower_bound}) holds iff
\begin{equation}
\label{eq:regret_equiv}
\frac{\frac{1}{r}\left(\frac{r}{2} - \min_{i}{A_{r,i}}\right)}{\frac{1}{r}\sqrt{r/2\ln{n}}} = \frac{\frac{r}{2} - \min_{i}{A_{r,i}}}{\sqrt{r/2\ln{n}}} \geq 1
\end{equation}
for $r$ and $n$ sufficiently large.


   In order to prove (\ref{eq:regret_equiv}), define $B_{r,n}$ by multiplying the left-hand side of (\ref{eq:regret_equiv}) by $-1$, so that
  \begin{displaymath}
  B_{r,n} \triangleq \frac{\min_{i}{A_{r,i}} - \frac{r}{2}}{\sqrt{r/2\ln{n}}}.
  \end{displaymath}
  Proving Equation (\ref{eq:regret_equiv}) is then equivalent to showing that, for $r$ and $n$ sufficiently large and any $\kappa > 0$, $\mathbf{E}(B_{r,n}) \leq -1 + \kappa$.  For any random variable $X$, we have that
  \begin{IEEEeqnarray*}{rCl}
  \mathbf{E}(X) &=& \mathbf{E}(X | X \leq -1 + \frac{\kappa}{3})Pr\left(X \leq -1 + \frac{\kappa}{3}\right) + \mathbf{E}(X | X \in [-1 + \frac{\kappa}{3}, 0])Pr(X \in [-1 + \kappa, 0]) + \\
  && \mathbf{E}(X | X \geq 0)Pr(X \geq 0) \\
  &\leq& \left(-1 + \frac{\kappa}{3}\right)Pr\left(X \leq -1 + \frac{\kappa}{3}\right) + 0\cdot Pr\left(X \in [-1 + \frac{\kappa}{3},0]\right) + \mathbf{E}(X | X \geq 0)Pr(X \geq 0).
  \end{IEEEeqnarray*}
  Hence for $B_{r,n}$ in particular we have
  \begin{equation}
  \label{eq:unknown_dynamic_upper_bound}
  \mathbf{E}(B_{r,n}) \leq \left(-1 + \frac{\kappa}{3}\right)Pr\left(B_{r,n} \leq -1 + \frac{\kappa}{3}\right) + \int_{0}^{\infty}{Pr(B_{r,n} \geq c) \ dc}.
  \end{equation}
 First, note that
  \begin{IEEEeqnarray}{rCl}
  \nonumber
  Pr(B_{r,n} \geq c) &=& Pr(A_{r,1} - r/2 \geq c\sqrt{r(\ln{n})/2}, \ldots, A_{r,n} - r/2 \geq c\sqrt{r(\ln{n})/2}) \\
  \label{eq:regret1}
   &=& \prod_{i=1}^{n}{Pr(A_{r,i} \geq r/2 + c \sqrt{r(\ln{n})/2})} \\
  \label{eq:regret2}
   &=& \left(Pr\left(A_{r,1} \geq \frac{r}{2} + r\left(\frac{c\sqrt{(\ln{n})/2}}{\sqrt{n}}\right)\right)\right)^{n} \\
   \label{eq:regret3}
   &\leq& \left(\exp{\left(-2\left(c\sqrt{\frac{2\ln{n}}{r}}\right)^{2}r\right)}\right)^{n}
   = \exp(-4c^{2}n\ln{n}),
   \end{IEEEeqnarray}
   where (\ref{eq:regret1}) and (\ref{eq:regret2}) follow from the fact that $(A_{r,1},\ldots,A_{r,n})$ are i.i.d. random variables and (\ref{eq:regret3}) follows from Hofferding's inequality.
   Hence
   \begin{displaymath}
   \int_{0}^{\infty}{Pr(B_{r,n} \geq c) \ dc} \leq \int_{0}^{\infty}{\exp{(-4c^{2}n\ln{n})} \ dc} = \frac{1}{8}\sqrt{\frac{\pi}{n\ln{n}}}
   \end{displaymath}
   which is less than $\epsilon/3$ for $n$ sufficiently large.

   Examining the first term of (\ref{eq:unknown_dynamic_upper_bound}), we first define $Y_{r,i,n} = \frac{A_{r,i} - \frac{r}{2}}{\sqrt{r}/2}$.  By the Central Limit Theorem and the fact that $A_{r,i}$ is binomial, $Y_{r,i,n}$ converges to a $N(0,1)$ random variable as $r \rightarrow \infty$.  Hence
   \begin{IEEEeqnarray*}{rCl}
   Pr\left(B_{r,n} \leq -1 + \frac{\kappa}{3}\right) &=& Pr\left(\min_{1 \leq i \leq n}{Y_{r,i,n}} \leq -1 + \frac{\kappa}{3}\right)
   = 1 - Pr\left(\min_{1 \leq i \leq n}{Y_{r,i,n}} \geq -1 + \frac{\kappa}{3}\right) \\
   &=& 1 - Pr\left(Y_{r,1,n} \geq -1 + \frac{\kappa}{3}\right)^{n}
   \geq 1 - \frac{\kappa}{3}
   \end{IEEEeqnarray*}
   for $n$ sufficiently large.  We therefore have
   \begin{displaymath}
   (-1 + \kappa/3)Pr(B_{r,n} \leq -1 + \kappa/3) \leq \left(-1 + \frac{\kappa}{3}\right)\left(1 - \frac{\kappa}{3}\right) < -1 + \frac{2\kappa}{3},
   \end{displaymath}
   and hence, (\ref{eq:unknown_dynamic_upper_bound}) reduces to
   $\mathbf{E}(B_{r,n}) \leq -1 + \frac{2\kappa}{3} + \frac{\kappa}{3} < -1 + \kappa,$
  which yields (\ref{eq:regret_equiv}), thus proving the theorem.
  \end{IEEEproof} 

%% file: Convergence_error_ArXiv.bbl
\begin{thebibliography}{10}
\providecommand{\url}[1]{#1}
\csname url@samestyle\endcsname
\providecommand{\newblock}{\relax}
\providecommand{\bibinfo}[2]{#2}
\providecommand{\BIBentrySTDinterwordspacing}{\spaceskip=0pt\relax}
\providecommand{\BIBentryALTinterwordstretchfactor}{4}
\providecommand{\BIBentryALTinterwordspacing}{\spaceskip=\fontdimen2\font plus
\BIBentryALTinterwordstretchfactor\fontdimen3\font minus
  \fontdimen4\font\relax}
\providecommand{\BIBforeignlanguage}[2]{{%
\expandafter\ifx\csname l@#1\endcsname\relax
\typeout{** WARNING: IEEEtran.bst: No hyphenation pattern has been}%
\typeout{** loaded for the language `#1'. Using the pattern for}%
\typeout{** the default language instead.}%
\else
\language=\csname l@#1\endcsname
\fi
#2}}
\providecommand{\BIBdecl}{\relax}
\BIBdecl

\bibitem{ren2005survey}
W.~Ren, R.~W.~Beard, and E.~M.~Atkins, ``A survey of consensus problems in
  multi-agent coordination,'' \emph{Proceedings of the American Control
  Conference}, pp. 1859--1864, 2005.

\bibitem{olfati2006flocking}
R.~Olfati-Saber, ``Flocking for multi-agent dynamic systems: Algorithms and
  theory,'' \emph{IEEE Transactions on Automatic Control}, vol.~51, no.~3, pp.
  401--420, 2006.

\bibitem{lawton2003decentralized}
J.~R.~T.~Lawton, R.~W.~Beard, and B.~J.~Young, ``A decentralized approach to
  formation maneuvers,'' \emph{IEEE Transactions on Robotics and Automation},
  vol.~19, no.~6, pp. 933--941, 2003.

\bibitem{olfati2009kalman}
R.~Olfati-Saber, ``Kalman-consensus filter: Optimality, stability, and
  performance,'' \emph{Proceedings of the 48th IEEE Conference on Decision and
  Control}, pp. 7036--7042, 2009.

\bibitem{ghaderi2012opinion}
J.~Ghaderi and R.~Srikant, ``Opinion dynamics in social networks: A local
  interaction game with stubborn agents,'' \emph{arXiv preprint
  arXiv:1208.5076}, 2012.

\bibitem{swaroop1996string}
D.~Swaroop and J.~Hedrick, ``String stability of interconnected systems,''
  \emph{IEEE Transactions on Automatic Control}, vol.~41, no.~3, pp. 349--357,
  1996.

\bibitem{degroot1974reaching}
M.~H.~DeGroot, ``Reaching a consensus,'' \emph{Journal of the American
  Statistical Association}, vol.~69, no. 345, pp. 118--121, 1974.

\bibitem{ren2007distributed}
W.~Ren and R.~W.~Beard, \emph{{Distributed Consensus in Multi-vehicle
  Cooperative Control: Theory and Applications}}.\hskip 1em plus 0.5em minus
  0.4em\relax Springer, 2007.

\bibitem{shah2009gossip}
D.~Shah, \emph{{Gossip Algorithms}}.\hskip 1em plus 0.5em minus 0.4em\relax Now
  Publishers, 2009.

\bibitem{yuan2012decentralised}
Y.~Yuan, J.~Liu, R.~M.~Murray, and J.~Gon{\c{c}}alves, ``Decentralised
  minimal-time dynamic consensus,'' \emph{Proceedings of the American Control
  Conference}, pp. 800--805, 2012.

\bibitem{boyd2006randomized}
S.~Boyd, A.~Ghosh, B.~Prabhakar, and D.~Shah, ``Randomized gossip algorithms,''
  \emph{IEEE Transactions on Information Theory}, vol.~52, no.~6, pp.
  2508--2530, 2006.

\bibitem{jadbabaie2003coordination}
A.~Jadbabaie, J.~Lin, and A.~S. Morse, ``Coordination of groups of mobile
  autonomous agents using nearest neighbor rules,'' \emph{IEEE Transactions on
  Automatic Control}, vol.~48, no.~6, pp. 988--1001, 2003.

\bibitem{tanner2004controllability}
H.~Tanner, ``On the controllability of nearest neighbor interconnections,''
  \emph{43rd IEEE Conference on Decision and Control (CDC)}, vol.~3, pp.
  2467--2472, 2004.

\bibitem{liu2008controllability}
B.~Liu, T.~Chu, L.~Wang, and G.~Xie, ``Controllability of a leader--follower
  dynamic network with switching topology,'' \emph{IEEE Transactions on
  Automatic Control}, vol.~53, no.~4, pp. 1009--1013, 2008.

\bibitem{olshevsky2009convergence}
A.~Olshevsky and J.~N.~Tsitsiklis, ``Convergence speed in distributed consensus
  and averaging,'' \emph{SIAM Journal on Control and Optimization}, vol.~48,
  no.~1, pp. 33--55, 2009.

\bibitem{rahmani2009controllability}
A.~Rahmani, M.~Ji, M.~Mesbahi, and M.~Egerstedt, ``Controllability of
  multi-agent systems from a graph-theoretic perspective,'' \emph{SIAM Journal
  on Control and Optimization}, vol.~48, no.~1, pp. 162--186, 2009.

\bibitem{pasqualetti2008steering}
F.~Pasqualetti, S.~Martini, and A.~Bicchi, ``Steering a leader-follower team
  via linear consensus,'' \emph{Hybrid Systems: Computation and Control}, pp.
  642--645, 2008.

\bibitem{mesbahi2010graph}
M.~Mesbahi and M.~Egerstedt, \emph{Graph {T}heoretic {M}ethods in {M}ultiagent
  {N}etworks}.\hskip 1em plus 0.5em minus 0.4em\relax Princeton University
  Press, 2010.

\bibitem{cao2012overview}
Y.~Cao, W.~Yu, W.~Ren, and G.~Chen, ``An overview of recent progress in the
  study of distributed multi-agent coordination,'' \emph{To appear in IEEE
  Transactions on Industrial Informatics}, 2012.

\bibitem{patterson2010convergence}
S.~Patterson, B.~Bamieh, and A.~El~Abbadi, ``Convergence rates of distributed
  average consensus with stochastic link failures,'' \emph{IEEE Transactions on
  Automatic Control}, vol.~55, no.~4, pp. 880--892, 2010.

\bibitem{cai2012convergence}
K.~Cai and H.~Ishii, ``Convergence time analysis of quantized gossip consensus
  on digraphs,'' \emph{Automatica}, vol.~48, no.~9, pp. 2344--2351, 2012.

\bibitem{ji2008containment}
M.~Ji, G.~Ferrari-Trecate, M.~Egerstedt, and A.~Buffa, ``Containment control in
  mobile networks,'' \emph{IEEE Transactions on Automatic Control}, vol.~53,
  no.~8, pp. 1972--1975, 2008.

\bibitem{notarstefano2011containment}
G.~Notarstefano, M.~Egerstedt, and M.~Haque, ``Containment in leader--follower
  networks with switching communication topologies,'' \emph{Automatica},
  vol.~47, no.~5, pp. 1035--1040, 2011.

\bibitem{ji2006leader}
M.~Ji, A.~Muhammad, and M.~Egerstedt, ``Leader-based multi-agent coordination:
  Controllability and optimal control,'' \emph{American Control Conference
  (ACC)}, pp. 1358--1363, 2006.

\bibitem{liu2011controllability}
Y.~Liu, J.~Slotine, and A.~Barab{\'a}si, ``Controllability of complex
  networks,'' \emph{Nature}, vol. 473, no. 7346, pp. 167--173, 2011.

\bibitem{patterson2010leader}
S.~Patterson and B.~Bamieh, ``Leader selection for optimal network coherence,''
  in \emph{Proceedings of the 49th IEEE Conference on Decision and Control
  (CDC)}.\hskip 1em plus 0.5em minus 0.4em\relax IEEE, 2010, pp. 2692--2697.

\bibitem{fardad2011noisefree}
M.~Fardad, F.~Lin, and M.~Jovanovic, ``Algorithms for leader selection in large
  dynamical networks: Noise-free leaders,'' \emph{50th IEEE Conference on
  Decision and Control and European Control Conference (CDC-ECC)}, pp.
  7188--7193, 2011.

\bibitem{lin2011noisecorrupted}
F.~Lin, M.~Fardad, and M.~Jovanovic, ``Algorithms for leader selection in large
  dynamical networks: Noise-corrupted leaders,'' in \emph{Proceedings of the
  50th IEEE Conference on Decision and Control and European Control Conference
  (CDC-ECC)}.\hskip 1em plus 0.5em minus 0.4em\relax IEEE, 2011.

\bibitem{kawashima2012leader}
H.~Kawashima and M.~Egerstedt, ``Leader selection via the manipulability of
  leader-follower networks,'' \emph{Proceedings of the American Control
  Conference}, pp. 6053--6058, 2012.

\bibitem{clark2012supermodular}
A.~Clark, L.~Bushnell, and R.~Poovendran, ``A supermodular optimization
  framework for leader selection under link noise in linear multi-agent
  systems,'' \emph{arXiv preprint arXiv:1208.0946}, 2012.

\bibitem{clark2012controllability}
------, ``On leader selection for performance and controllability in
  multi-agent systems,'' \emph{51st IEEE Conference on Decision and Control
  (CDC)}, pp. 86--93, 2012.

\bibitem{clark2012leader}
------, ``Leader selection for minimizing convergence error in leader-follower
  systems: A supermodular optimization approach,'' \emph{10th International
  Symposium on Modeling and Optimization in Mobile, Ad Hoc and Wireless
  Networks (WiOpt)}, pp. 111--115, 2012.

\bibitem{kempe2003maximizing}
D.~Kempe, J.~Kleinberg, and {\'E}.~Tardos, ``Maximizing the spread of influence
  through a social network,'' \emph{{Proceedings of the 9th ACM SIGKDD
  International Conference on Knowledge Discovery and Data Mining}}, pp.
  137--146, 2003.

\bibitem{borkar2010manufacturing}
V.~S.~Borkar, J.~Nair, and N.~Sanketh, ``Manufacturing consent,''
  \emph{Proceedings of the 48th Annual Allerton Conference on Communication,
  Control, and Computing}, pp. 1550--1555, 2010.

\bibitem{wolsey1999integer}
L.~Wolsey and G.~Nemhauser, \emph{Integer and {C}ombinatorial
  {O}ptimization}.\hskip 1em plus 0.5em minus 0.4em\relax Wiley-Interscience,
  1999.

\bibitem{fujishige2005submodular}
S.~Fujishige, \emph{{Submodular Functions and Optimization}}.\hskip 1em plus
  0.5em minus 0.4em\relax Elsevier Science, 2005.

\bibitem{chatterjee1977towards}
S.~Chatterjee and E.~Seneta, ``Towards consensus: Some convergence theorems on
  repeated averaging,'' \emph{Journal of Applied Probability}, pp. 89--97,
  1977.

\bibitem{tahbaz2008necessary}
A.~Tahbaz-Salehi and A.~Jadbabaie, ``A necessary and sufficient condition for
  consensus over random networks,'' \emph{IEEE Transactions on Automatic
  Control}, vol.~53, no.~3, pp. 791--795, 2008.

\bibitem{cao2008reaching}
M.~Cao, A.~S.~Morse, and B.~D.~O.~Anderson, ``Reaching a consensus in a
  dynamically changing environment: A graphical approach,'' \emph{SIAM Journal
  on Control and Optimization}, vol.~47, no.~2, pp. 575--600, 2008.

\bibitem{jakovetic2010weight}
D.~Jakovetic, J.~Xavier, and J.~M.~F.~Moura, ``Weight optimization for
  consensus algorithms with correlated switching topology,'' \emph{IEEE
  Transactions on Signal Processing}, vol.~58, no.~7, pp. 3788--3801, 2010.

\bibitem{bettstetter2003node}
C.~Bettstetter, G.~Resta, and P.~Santi, ``The node distribution of the random
  waypoint mobility model for wireless ad hoc networks,'' \emph{IEEE
  Transactions on Mobile Computing}, vol.~2, no.~3, pp. 257--269, 2003.

\bibitem{clark2013arxiv}
A.~Clark, B.~Alomair, L.~Bushnell, and R.~Poovendran, ``Minimizing convergence
  error in multi-agent systems via leader selection: A supermodular
  optimization approach,'' \emph{arXiv preprint arXiv:1306.4949}, 2013.

\bibitem{cesa2006prediction}
N.~Cesa-Bianchi and G.~Lugosi, \emph{{P}rediction, {L}earning, and
  {G}ames}.\hskip 1em plus 0.5em minus 0.4em\relax {C}ambridge {U}niversity
  {P}ress, 2006.

\bibitem{streeter2007online}
M.~Streeter and D.~Golovin, ``{An online algorithm for maximizing submodular
  functions},'' in \emph{Carnegie Mellon University Technical Report}, 2007.

\bibitem{freund1995desicion}
Y.~Freund and R.~E.~Schapire, ``A desicion-theoretic generalization of on-line
  learning and an application to boosting,'' \emph{Computational Learning
  Theory}, pp. 23--37, 1995.

\bibitem{hong1999group}
X.~Hong, M.~Gerla, G.~Pei, and C.~Chiang, ``A group mobility model for ad hoc
  wireless networks,'' \emph{2nd ACM international workshop on Modeling,
  analysis and simulation of wireless and mobile systems}, pp. 53--60, 1999.

\end{thebibliography}
